%% file: main.tex
\pdfoutput=1
\documentclass[11pt]{article}

\usepackage[]{ACL2023}

\usepackage{times}
\usepackage{latexsym}

\usepackage[T1]{fontenc}
\usepackage[utf8]{inputenc}

\usepackage{microtype}

\usepackage{inconsolata}

\usepackage{times}
\usepackage{latexsym}
\usepackage{graphicx}
\usepackage{xcolor}
\usepackage{colortbl}
\usepackage{amsmath}
\usepackage{tcolorbox}
\usepackage{booktabs}
\usepackage{array}
\usepackage{multirow}
\usepackage{subcaption}
\usepackage{setspace}
\usepackage{multicol}
\usepackage{tcolorbox}
\usepackage{multicol}
\usepackage{hyperref}
\tcbuselibrary{skins, breakable}
\usepackage{bbding}
\usepackage{xspace}
\usepackage{algpseudocode}
\usepackage{algorithm}
\usepackage{fancyvrb}
\usepackage[framemethod=default]{mdframed}
\definecolor{mygray}{gray}{0.6}
\definecolor{mygray-bg}{gray}{0.95}
\usepackage{float}

\newcommand{\dataset}{\texttt{MusicPile}\xspace}

\tcbset{
  before skip=5pt,
  after skip=5pt
}

\newtcolorbox{userbox}{
  breakable,
  colback=blue!5!white,
  colframe=blue!75!black,
  title=User Proxy Agent,
  fonttitle=\bfseries
}

\newtcolorbox{arbox}{
  breakable,
  colback=green!5!white,
  colframe=green!75!black,
  title=A\&R Agent,
  fonttitle=\bfseries
}

\newtcolorbox{melodybox}{
  breakable,
  colback=orange!5!white,
  colframe=orange!75!black,
  title=Melody Agent,
  fonttitle=\bfseries
}

\newtcolorbox{harmonybox}{
  breakable,
  colback=red!5!white,
  colframe=red!75!black,
  title=Harmony Agent,
  fonttitle=\bfseries
}

\newtcolorbox{instrumentbox}{
  breakable,
  colback=magenta!5!white,
  colframe=magenta!75!black,
  title=Instrument Agent,
  fonttitle=\bfseries
}

\newtcolorbox{reviewertbox}{
  breakable,
  colback=lime!5!white,
  colframe=lime!75!black,
  title=Reviewer Agent,
  fonttitle=\bfseries
}

\title{ChatMusician: Understanding and Generating Music\\ Intrinsically with LLM}

\newcommand{\maplogo}{
    \includegraphics[scale=0.07]{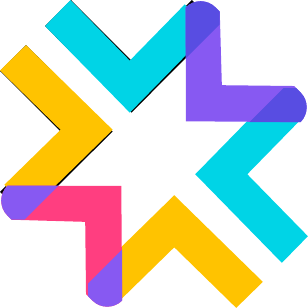}
}

\author{\textnormal{\maplogo Multimodal Art Projection Research Community} \\
\textnormal{Skywork AI PTE. LTD.} \\
\textnormal{Hong Kong University of Science and Technology}}

\begin{document}

\maketitle
\input{sections/0abstract}
% ruibin
\input{sections/1introduction}

% ruibin
\input{sections/2related_work}
% zhangge

\input{sections/3methods}

% pt data zeyue
% sft data hanfeng
    % music score sft data curation wangyi

\input{sections/5experiment}
% data mixture hanfeng
% training hanfeng

\input{sections/6results}

% subjective eval wangyi, ziyu
% objective eval tianhao, liumeng

\input{sections/7conclusion}

% ruibin

\input{sections/8limitations}

% ruibin

\section{Contributions and Acknowledgments}
\label{sec:9acknowledgement}
\input{sections/9acknowledgement}

% thanks kunlun

\bibliography{custom, anthology}
\bibliographystyle{acl_natbib}

\clearpage 
\input{sections/appendix}
% case study

\end{document}

%% file: sections/0abstract.tex
\begin{abstract}
% contributions  

% 1. the first study of injecting instrinsic musical abilities into LLM.

% 2. we open source code, data, model, benchmark.

% 3. our model writes music better than gpt4 and baselines

% 4. we contribute the first college-level symbolic music understanding benchmark. The benchmark contains music understanding and reasoning, and we find  that LLMs perform badly on the benchmark. It suggests that music is another land to conquer, besides code and math.

% 5. we unified multiple symbolic music understanding and generation tasks into one LLMs, while keeping or even enhancing the backbone's general abilities. LLM can serve as a music generalist. The creative side of LLMs can be extended to music.

% While Large Language Models (LLMs) demonstrate impressive capabilities in musical knowledge, we find that music reasoning\footnote{The ability to estimate the varying harmonies, keys, rhythms, and other musical elements that are not explicitly annotated in a piece of music and are significant for music themes, progression, and styles is called \textbf{Music Reasoning}.} is still an unsolved task.
While Large Language Models (LLMs) demonstrate impressive capabilities in text generation, we find that their ability has yet to be generalized to music, humanity’s creative language.
% with ChatMusician, or even GPT-4 only surpasses random baseline by a small margin.
We introduce \textbf{ChatMusician}\footnote{See Contributions and Acknowledgments section for full author list.}, an open-source LLM that integrates intrinsic musical abilities. 
It is based on continual pre-training and finetuning LLaMA2 on a text-compatible music representation, ABC notation, and the music is treated as a second language.
ChatMusician can understand and generate music with a pure text tokenizer without any external multi-modal neural structures or tokenizers. 
Interestingly, endowing musical abilities does not harm language abilities, even achieving a slightly higher MMLU score.
Our model is capable of composing well-structured, full-length music, conditioned on texts, chords, melodies, motifs, musical forms, etc, surpassing GPT-4 baseline.
On our meticulously curated college-level music understanding benchmark, \textbf{MusicTheoryBench}, ChatMusician surpasses LLaMA2 and GPT-3.5 on zero-shot setting by a noticeable margin. 
Our work reveals that LLMs can be an excellent compressor for music, but there remains significant territory to be conquered.
We release our 4B token music-language corpora \textbf{MusicPile}, the collected MusicTheoryBench, code, model and demo in \href{https://shanghaicannon.github.io/ChatMusician/}{GitHub}.

\end{abstract}

%% file: sections/1introduction.tex
\section{Introduction}
% background
% why ai4art, why music
    % art is important, because
        % creativity of humanity
    % music is important, because
        % similarity between symbolic music and language
The fusion of artificial intelligence and the arts, particularly music, has emerged as a pivotal area of research, for its profound implications on the essence of human creativity ~\citep{civit2022systematic}. 
Music holds a unique position due to its inherent structure and complexity, 
 and ~\citet{masataka2009origins, masataka2007music, pino2023association} suggest that language and music may have evolved from the same source. 
% often drawing parallels to the symbolic nature of language itself. 
\begin{figure*}[hbt]
\vspace{0.1cm}
\centering
\setlength{\abovecaptionskip}{0.1cm} 
\includegraphics[width=2.0\columnwidth]{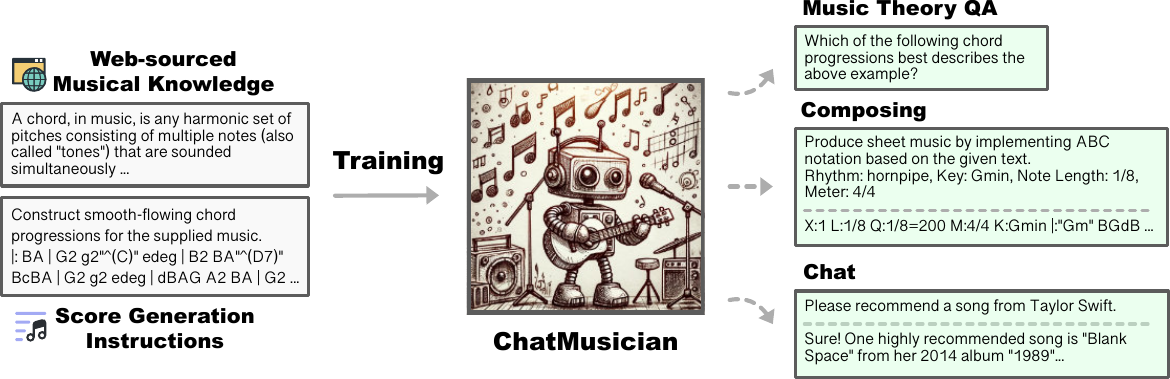}
\caption{ChatMusician learns from web-sourced musical knowledge and handcrafted music score generation instructions, unifies music generation and music understanding, and can chat, compose, and answer college-level music theory questions.}
\label{fig:pipeline}
\end{figure*}

Large Language Models (LLMs) have recently revolutionized various domains with their remarkable capacity for generating long sequences. 
Researchers have been exploring language modeling techniques for music generation~\citep{vaswani2017attention,huang2018musictransformer,payne2019musenet,lu2023musecoco,dhariwal2020jukebox,agostinelli2023musiclm,copet2023simple,margulis2016repetition,dai2022missing,jhamtani2019modeling}. 
Although it seems that symbolic music can be treated in a similar way to the natural language, these practices have shown that many distinct challenges are encountered when it comes to the realm of music. For example, even state-of-the-art models such as GPT-4 perform marginally better than random in music reasoning\footnote{The ability to estimate the varying harmonies, keys, rhythms, and other musical elements that are not explicitly annotated in a piece of music and are significant for music themes, progression, and styles is called \textbf{Music Reasoning}.}. We argue that the main reason is that the intricacies of musical composition remain inadequately represented in current LLMs, including the long-term, contrapuntal context dependency and the complex connections between music notes and text descriptions. 
Attempting to find solutions to these challenges, we propose ChatMusician, an open-source LLM that integrates intrinsic musical abilities, with pipeline as shown in Figure \ref{fig:pipeline}. Our endeavors have focused on leveraging LLMs for symbolic music generation and understanding. 

% what have we learned
    % music understanding, reasoning is non-trivial for llms, even gpt4 is near random. music is another field that may deserve research attention, both those studying reasoning, and generalization
    % copying?
% The fusion of artificial intelligence and the arts, particularly in the realm of music, has sparked significant interest owing to its implications for human creativity \citep{civit2022systematic}. 
% Music, sharing inherent structural parallels with language, stands as a unique domain where language models have been applied, given the theoretical connections between the two \citep{masataka2009origins, masataka2007music, pino2023association}. 
% Despite the success of language models in various tasks, the complexities inherent in music pose distinct challenges. 

% contributions
\textbf{Our contributions:}
% This includes a comprehensive five-fold approach. 
a) We introduce \textit{ChatMusician}, a text-based LLM that unifies multiple symbolic music understanding and generation tasks, enriching their repertoire while maintaining or potentially enhancing their foundational general abilities.
b)  Empirical evaluations demonstrate our model's superior musical composition capabilities, surpassing GPT-4 and established baselines in various music generation tasks, showcasing its prowess in generating coherent and structured musical pieces across diverse styles. 
c) We introduce the inaugural college-level symbolic music understanding benchmark, \textit{MusicTheoryBench}, comprising facets of music understanding and reasoning. LLMs' performance on this benchmark exposes their limitations, suggesting the uncharted territory of music as a domain demanding attention akin to code and mathematical reasoning. 
d) We open source the complete framework, including benchmark, codes, and 4B-token music-language corpora \textit{MusicPile}, fostering collaboration in this field.

%% file: sections/2related_work.tex
\section{Related Work}

\subsection{Issues in Music Generation and Understanding}
% music generation就是引用一下introduction里提到的work就行，jukebox，musecoco，musenet，music transformer，musiclm，musicgen。有audio的有symbolic的。

% Various setups have long been employed in the study of music generation ~\cite{dhariwal2020jukebox, lu2023musecoco}. For acoustic music generation, 
% Jukebox~\citep{dhariwal2020jukebox} proposes a model that generates music with singing conditioned on artist and genre to steer the musical and vocal style.
% MusicLM~\citep{agostinelli2023musiclm} approaches conditional audio music generation as a hierarchical sequence-to-sequence task.
% % producing consistent music at 24 kHz over extended durations.
% MusicGen~\citep{copet2023simple} can produce high-quality, mono and stereo samples, conditioned on text or melodic features for refined control.
% % 还要引用一下shangda和bob的工作。
% % not sure which paper you are referring to besides the following two. -- YM.
% For symbolic music generation, representative works include~\citep{sturm2015folk,lu2023musecoco,MuseNet,huang2018musictransformer,zhuo2023video,wu2022exploring}.
The study of \textbf{music generation} is divided into acoustic ~\citep{dhariwal2020jukebox, agostinelli2023musiclm, copet2023simple} and symbolic modality ~\citep{sturm2015folk,lu2023musecoco,MuseNet,huang2018musictransformer,zhuo2023video,wu2022exploring}. However, works generated by these models are still limited to a short context (like 30s in audio form) and are far away from being completely musical and well-structured. ~\citet{margulis2016repetition} claims that "repetition" has a significantly positive effect on how listeners rate the "musicality" of an excerpt even if it is a random sequence. Early rule-based methods realize repetition with some pre-defined patterns which lack flexibility, whereas ~\citet{dai2022missing} reveals that deep-learning-based works may lack repetition and music structure in the generated music.

The landscape of \textbf{music understanding} has traditionally centered on audio-focused tasks, exemplified by significant endeavors like the Music Information Retrieval Exchange (MIREX)\footnote{\scriptsize{\url{https://www.music-ir.org/mirex/wiki/MIREX_HOME}}} data challenge and the MARBLE benchmark~\citep{yuan2023marble}. For instance, they tackled various audio-based tasks such as genre classification, chord estimation, melody extraction, etc. In contrast, our contribution stands out with the introduction of the MusicTheoryBench, diverging from the conventional audio-centric focus by encompassing challenges in music verbal comprehension, advanced music theory understanding and symbolic music reasoning. 
% This benchmark presents a unique set of challenges, probing the depths of symbolic music interpretation and theoretical understanding, providing a robust evaluation platform for the nuanced aspects of music comprehension beyond audio, thus contributing significantly to the domain of music information retrieval.

% \begin{figure*}[t]
% \centering
% \includegraphics[width=\textwidth]{figures/representation.pdf}
% \caption[]{Commonly used music representations, including Wav, Codec, MIDI and ABC notation. From left to right, the compression rate gets higher.
% } 
% \label{fig:representation}
% \end{figure*}

\subsection{Music Representations}

\begin{figure*}[hbt]
\centering
\includegraphics[width=\textwidth]{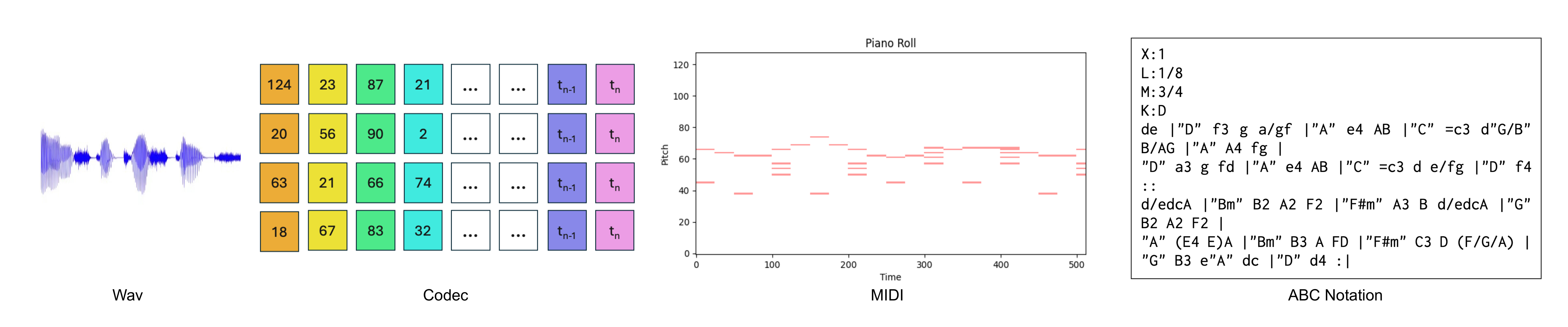}
\caption[]{Commonly used music representations, including Wav, Codec, MIDI (visualized as piano roll), and ABC notation. From left to right, the compression rate gets higher.
} 
\label{fig:representation}
\end{figure*}

% 给出music information hiearchy （问ruibin），贴一个图然后解释一下
% 然后说我们在讨论symbolic music，比起audio和codec，和language更接近
% 音乐某种程度上来说是一种高级语言吗，近年来natrural language processing领域取得了卓越的发展，一些音频上的工作也尝试用codec模型把audio feature进行token化，把audio当作一种"语言"，从而实现用nlp的技术来训练audio(Soundstream, encodec)模型。相较于codec, symbolic music representation更接近于一种语言. Symbolic music是一种类似于谱子,将音乐表达为symbolic data的形式.
% Music, to an extent, can be viewed as a sophisticated language variant. 
% The progression in Natural Language Processing (NLP) has initiated novel methods for handling audio features via codec models, translating audio into a language-like format for audio model training. For examples, MusicLM~\citep{agostinelli2023musiclm} uses Soundstream~\citep{DBLP:journals/taslp/ZeghidourLOST22}, while MusicGen~\citep{copet2023simple} adopts EnCodec~\citep{défossez2022high}. 
% In contrast, symbolic music representation, more closely resembling language, offers a different approach. 
% Symbolic music, either as musical scores or encoded performance data, represents music symbolically, differing from codec model techniques.

% symbolic music有几种常用的representation，比如performance midi，score midi，kern score，abc notation（link到appendix A）。
% 以往的work通常用performance midi，用REMI-like或MIDI-like的midi representation，
% 我们一般不用score midi，因为数据量很少很难获取。一般用performance midi，数据量多一些，有很多开源数据集。
Figure \ref{fig:representation} displays mainstream music representations with varying compression rates.
Symbolic music includes formats like MIDI, humdrum, and ABC notation (detailed in Appendix \ref{sec:abc_introducion}). 
MIDI has been a research favorite~\cite{lu2023musecoco,DBLP:conf/mm/HuangY20,DBLP:conf/iclr/HuangVUSHSDHDE19} with easily-accessible data due to its popularity in the music industry. 
However, to solve MIDI's lengthy sequence challenges posed for transformer models with intensive training demands, the sequences are typically segmented into shorter fragments which limit capturing a composition's full continuity. 
Additionally, MIDI's encoding of performance nuances can lead to quantization errors and unstable rhythms when being tokenized.
% , as seen in Tanxu's Musecoco, Anna's Music Transformer, and the REMI paper (Pop Music Transformer: Beat-based Modeling and Generation of Expressive Pop Piano Compositions). Score MIDI is infrequently used due to limited data availability, whereas performance MIDI, with larger available datasets, is more commonly adopted.
% 添加下列应用: musecoco，anna的music transformer，REMI paper（Pop Music Transformer: Beat-based Modeling and Generation of Expressive Pop Piano Compositions），

% performance midi representation的sequence length很长，通常需要切片成几十秒的片段来训练。几十秒就对应4k的token序列了，而transformer的attention的复杂度是O(n^2)的，复杂度很高（music transformer还特地提出了一种简化的attention机制，可以去看看他的讨论，转述一下）。
% 还有一个问题是，performance midi包含了演奏信息，然后做完tokenization quantize过后会有quantization error，听起来拍子很不稳定。

% 而我们这里使用的abc notation是score level的representation，有诸多好处：
% 0. 数据量大，irish musician大力发展这种representation，文档全面，而且网上谱子多
% 1. score没有quantization的问题，生成的东西都在拍子上
% 2. 压缩率很高，seq len短。typical长度大概300 token一首歌。利用谱面的反复记号自然的encode了music repetition和structure，降低计算复杂度。并且我们展示abc这种representation能自然的学到repetition和structure。
% 3. 对于llm而言，不需要对tokenizer进行额外的改动，是一种和字符串兼容的representation。
% 4. 人类友好，可懂性强。这也有利于基于LLM的music understanding和generation。
Therefore, we employ ABC notation, a score-oriented and plain text representation, for its notable advantages. 
% This score-oriented, plain text representation has extensive data availability, largely due to contributions from Irish musicians and online repositories. 
Its high compression rate leads to shorter sequence lengths compared to MIDI and it intrinsically encodes musical repetition and structure (e.g. by the use of repeated symbols), enhancing processing efficiency using language models.
It also includes detailed musical symbols denoting performance techniques and avoids quantization issues, ensuring rhythmic precision in music generation. 
ABC notation's compatibility with language models also facilitates its integration into LLM applications, allowing for advanced musical analysis and generation.

% \includegraphics[width=.8\textwidth]{figures/representation.pdf}
% \caption{ABC notation and corresponding staff notation of a generated music. Repetition symbols are marked blue in both notations and demonstrates a clear phrase level repetition. Red and yellow rectangles mark clear motif level repetition in both sections. Green rectangles mark variation notes following the motif of first section.}
% \label{fig:generation_sample}

% 贴一个各种symbolic music representation的可视化对比图

\subsection{LLMs for Complex Problem-solving Tasks in Non-language Domain}
% math, code, chessgpt, othellogpt
% \subsection{Bridging Language and Non-language Sequence Modeling}
% In the realm of bridging language and non-language sequence modeling, recent advancements have showcased the versatility and capability of LLMs in understanding and generating not just textual content but also engaging in complex decision-making and problem-solving tasks across various domains.
% ~\citet{li2022emergent} explore the application of a GPT variant to Othello, revealing the model's capacity to predict legal moves and suggesting emergent internal representations of non-linguistic data. This highlights LLMs' potential in understanding complex game dynamics without explicit prior knowledge.
% ~\citet{yue2023mammoth} present MAmmoTH, an LLM trained for solving mathematical problems, leveraging a hybrid approach of chain-of-thought and program-of-thought rationales. This underscores the models' ability to process and solve structured logical tasks, bridging language understanding with mathematical reasoning.
% ~\citet{feng2023chessgpt} introduce ChessGPT, which integrates historical chess game data and analytical insights in natural language, showcasing the fusion of policy learning with language modeling. This approach points towards creating more intuitive and strategic autonomous agents.
% ~\citet{roziere2023code} release Code Llama, a suite of LLMs for programming tasks, demonstrating proficiency in code generation and instruction following. This work exemplifies LLMs' capabilities in applying textual instructions to generate coherent and functional code sequences.

To well understand and generate music, a model needs to handle complex sequential modeling concerning motifs, harmonies, rhythms, texture, etc., compromising between a well-organized structure and divergent creativity. Based on the fundamental language sequence modeling, recent LLMs' advancements have showcased their generalization ability in complex decision-making and problem-solving tasks across non-language domains like maths, codes, and games, but have not considered music yet.
MAmmoTH~\citep{yue2023mammoth} leverage a hybrid approach of chain-of-thought and program-of-thought rationales to process and solve structured logical tasks, bridging language understanding with mathematical reasoning.
CodeLLaMA~\citep{roziere2023code}, a suite of LLMs for programming tasks, exemplifies LLMs' capabilities in applying textual instructions to generate coherent and functional code sequences.
Othello-GPT~\citep{li2022emergent} apply a variant of GPT, using nonlinear probe representations, layerwise interventions, and latent saliency maps, to predict legal moves in the Othello game.
ChessGPT~\citep{feng2023chessgpt} integrates historical chess game data and analytical insights in natural language, showcasing the fusion of policy learning with language modeling.

%% file: sections/3methods.tex
\section{Method}
\label{sec3}

\subsection{Language Corpora Curation}

% 体现多样性
% 体现comprehensive
\input{tables/pretrain_dataset}
% Ideally, we want to develop musical competence using musical knowledge from a natural corpus. 
To the best of our knowledge, there is currently no publicly available music-related natural language corpus. 
Fortunately, there are many large-scale corpora available from which we can curate our own. 
To enable our model to interact and conversationally receive instructions, we use data from various domains. 
In this section, we introduce our dataset \dataset, a first-of-its-kind pretraining dataset for injecting musical abilities into LLMs. 

\paragraph{General corpora.} Representative public datasets, including Pile \cite{gao2020pile}, Falcon RefinedWeb \cite{penedo2023refinedweb} and Wikipedia \cite{2023wiki} are used.
% The Pile is a large and diverse textual dataset, with much of its content sourced from academic or professional sources. 
% RefinedWeb is a massive English web dataset which is built through stringent filtering and large-scale deduplication of CommonCrawl. 
% Both of them have been properly preproccessed and used to train LLMs. 
To curate a musically relevant corpus, we list a set of music-related words as a criterion to filter Pile, based on music terminologies\footnote{\url{https://en.m.wikipedia.org/wiki/Glossary_of_music_terminology}}. 
% To ensure high relevance, we only include music terminology words that appear more than 10 times and account for over 0.5\% of the section. 
We only include music terminology words that appear more than 10 times and account for over 0.5\% of domain agreement. 
% In addition, we added the latest Wikipedia, where we hope the latest entries will augment the music-related knowledge base. 
% Wikipedia is a high-quality corpus and we want to maintain the generalization ability of the model, thus we just do simple filtering. 
%    no need for wordcloud 
% We generate a word cloud using the filtered results from these three corpora shown in Figure \ref{wordcloud}.
% \begin{figure}[hbt]
% \vspace{0.1cm}
% \centering
% \setlength{\abovecaptionskip}{0.1cm} 
% \includegraphics[width=1.0\columnwidth]{figures/pretrain_set_wordcloud.pdf}
% \caption{Wordcloud of filtered pile, refinedweb and wiki}
% \label{wordcloud}
% \end{figure}
%%%%%%%%%%%%%%%%%%%%%%%%
\paragraph{Instruction and chat data.} 
% \citet{DatabricksBlog2023DollyV2, peng2023instruction, wang2023far} have shown that fine-tuning LLMs is beneficial for generalizing models to new scenarios without training. 
The instruction datasets \citet{DatabricksBlog2023DollyV2, peng2023instruction, wang2023far} are diverse and representative enough to adapt the LLM to potential downstream usage. 
To enable multiple rounds of conversations, chat corpora \cite{wang2023openchat} are included. 

% \paragraph{Music scores} Symbolic music datasets are limited in the computational music community. To make full use of them, we have crafted a few more instructions on top of them. We hope to use this to help the model develop a strong symbolic music generation capability, which is expanded in detail in Section \ref{sec:music_score_set}. We have labeled a portion of the world map with the distribution of music scores containing regional information. As shown in Figure \ref{scatter}, our music scores is characterized by diversity.
\paragraph{Music knowledge and music summary.}  
We crawl the metadata corresponding to 2 million music tracks from YouTube, including metadata such as song title, description, album, artist, lyrics, playlist, etc. 
500k of them are extracted. 
We generate summaries of these metadata using GPT-4. 
We generate music knowledge QA pairs following Self-instruct\cite{wang2022self}. According to our topic outline in Appendix~\ref{ref:music knowledge}), 255k instructions are generated, with corresponding answers generated with GPT-4.
% music summary
%   1. 从youtube爬取了2M个music对应的metadata，包括：曲名、description、album、artist、lyrics、playlists等信息，抽取了其中的500k条
%   2. 送进GPT-3.5 summarize生成对应的music captions。（可以在appendix给一个instruction、input、output的例子）

% music knowledge
%   1. 利用我们题目大纲，根据self instruct的方式调用GPT-3.5根据大纲生成music相关的knowledge instruction 255k条
%   2. 调用GPT-3.5去回答这些instruction。

\paragraph{Math and code data.} 

The computational music community lacks symbolic music datasets, and we hypothesize that including math \cite{cobbe2021gsm8k, arxiv-math-instruct-50k, yue2023mammoth, li2023camel} and code \cite{li2023camel, wang2023openchat} may enhance the reasoning power of symbolic music. 
Empirically, we find this helps to improve the performance of music LLMs. 
% At worst, this addition serves to ensure the diversity of the dataset.

% \input{tables/pretrain_dataset}
% Except for the general corpora, all the other datasets were constructed as conversation forms, which we constructed as \textit{Human: \{content\} </s> Assistant: \{content\} </s> } for one or more rounds. 
% The percentage of musical verbal, code, music score, math, and general is 10.42\%, 2.43\%, 18.43\%, 4.05\%, and 64.68\%, respectively. 
% % shown in Figure \ref{tokens_pie}. 
% The overview of the pretrain set is in Table \ref{tab:pretrain_dataset}.
Except for the general corpora, all the other datasets were constructed as conversation forms for one or more rounds. 
The percentage of musical verbal, code, music score, math, and general is 10.42\%, 2.43\%, 18.43\%, 4.05\%, and 64.68\%, respectively. 
% shown in Figure \ref{tokens_pie}. 
Table~\ref{tab:pretrain_dataset} shows an overview of all data.

% \begin{figure}[hbt]
% \vspace{0.1cm}
% \centering
% \setlength{\abovecaptionskip}{0.1cm} 
% \includegraphics[width=1.0\columnwidth]{figures/tokens_pie_chart.pdf}
% \caption{Tokens proportion of \dataset}
% \label{tokens_pie}
% \end{figure}

\subsection{Music Score Corpora Curation}

\begin{figure}[hbt]
\vspace{0.1cm}
\centering
\setlength{\abovecaptionskip}{0.1cm} 
\includegraphics[width=1.0\columnwidth]{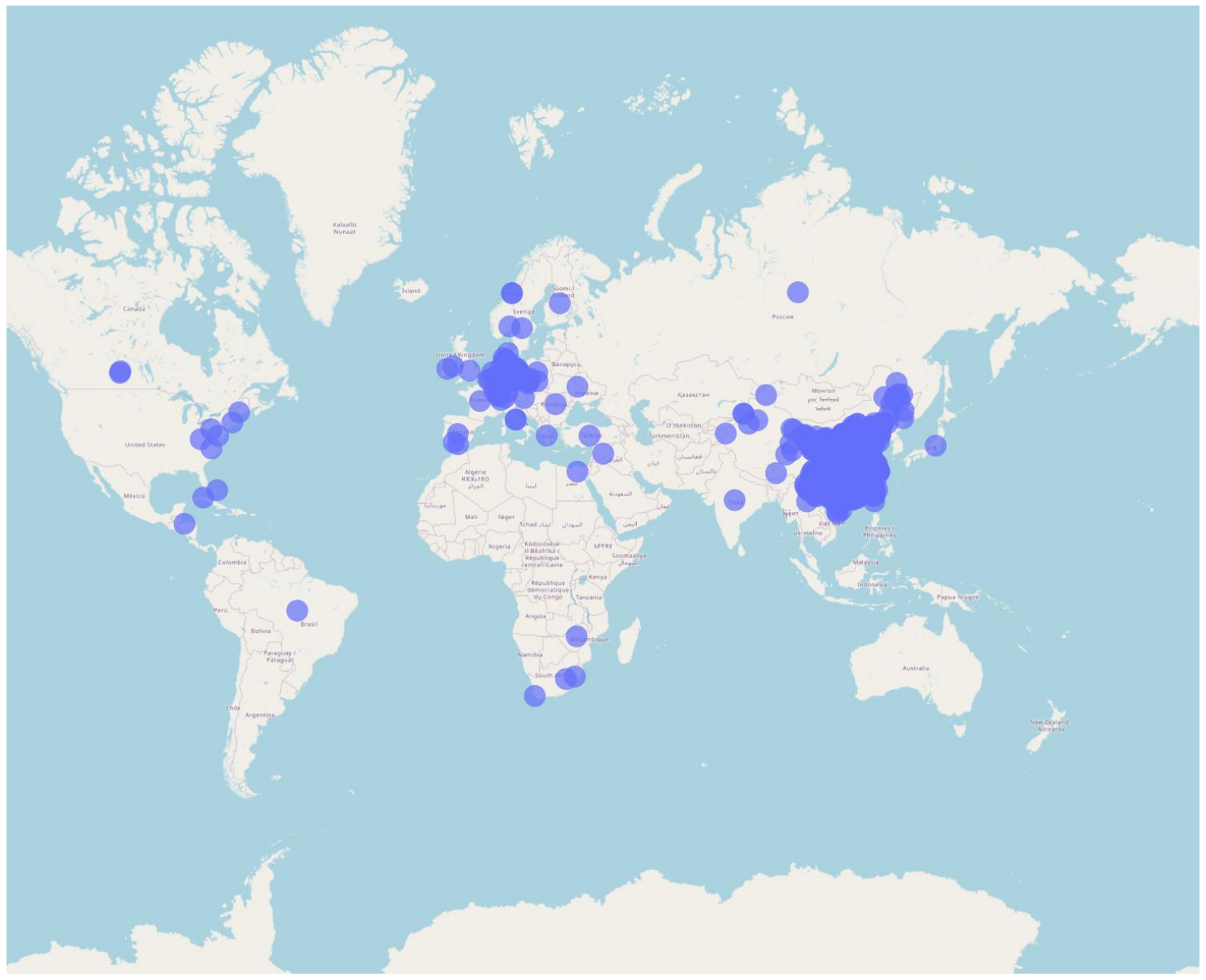}
\caption{We included diverse music scores from around the world in \dataset. The distribution of a portion of music scores containing regional information has been marked with blue points on the world map.}
\label{scatter}
\end{figure}

Although symbolic music datasets are scarce in the computational music community, we have made an effort to include music from various regions of the world. The distribution of a portion of music scores containing regional information has been labeled on the world map. As shown in Figure \ref{scatter}, our music scores are characterized by diversity. We designed a total of eight representative musical tasks on the collected corpora, including six for generating music scores and two for music understanding. 
The generative tasks involve generating music scores conditioned on the chord, melody, motif\footnote{In music, motif is a short musical idea, a salient recurring figure, musical fragment or succession of notes that has some special importance in or is characteristic of a composition}, musical form\footnote{In music, form refers to the structure of a musical composition or performance.}, and style.
The understanding tasks involve extracting motifs and forms from the user input scores.
For each task, we have created multiple instructions, which are listed in Table \ref{tab:tasks}, each with one example. The process of curating music instructions and algorithms is described in detail in Appendix \ref{sec:music_score_set}.

% Four tasks involve generating music scores conditioned on chord, music form, alphabetic musical form, or terminology musical form. 
% Musical forms are defined based on the similarity between sections in a composition, divided into three degrees: identical, variant, and different. 
% Alphabetic musical form and commonly used musical form terms are used to express the musical form of the compositions in the dataset. Instructions related to musical forms can be constructed from them. 

% Two of these tasks involve extracting motifs or musical forms from a piece of music. In music, a motif is a short musical idea, a salient recurring figure, a musical fragment, or a succession of notes that has some special importance in or is characteristic of a composition. An algorithm is designed to divide sections in a composition and extract the motif. Another task involves formulating chord combinations, and the last involves mimicking Bach's style to compose music. 

\input{tables/tasks}

\subsection{MusicTheoryBench}
\label{music_benchmark}
Despite the significant advancements in music information retrieval, the definition of advanced music understanding capabilities remains unclear in current research. In this study, to measure the advanced understanding abilities of existing LLMs in music, we first define two critical elements of music understanding: music knowledge and music reasoning. We then introduce MusicTheoryBench, a benchmark designed to assess the advanced music understanding capabilities of current LLMs.

\paragraph{Definition of Music Knowledge and Reasoning.} \textit{Reasoning} refers to the process of making inferences based on existing knowledge and observations, usually associated with math.\citep{yu2023nature, luo2023wizardmath}. Music is often likened to mathematics, where composers meticulously calculate the principles of form, harmony, scales, rhythm, tonality, and structural organization. This meticulous computation ensures that the distribution of notes across temporal and frequency domains meets established norms, yielding consonance and pleasing auditory experiences. The composition process frequently employs complex rules, including symmetry, transposition, repetition, inversion, and retrograde. 

\begin{figure}[ht]
\begin{mdframed}[innerleftmargin=1pt, innerrightmargin=1pt]
\footnotesize
\includegraphics[width=\linewidth]{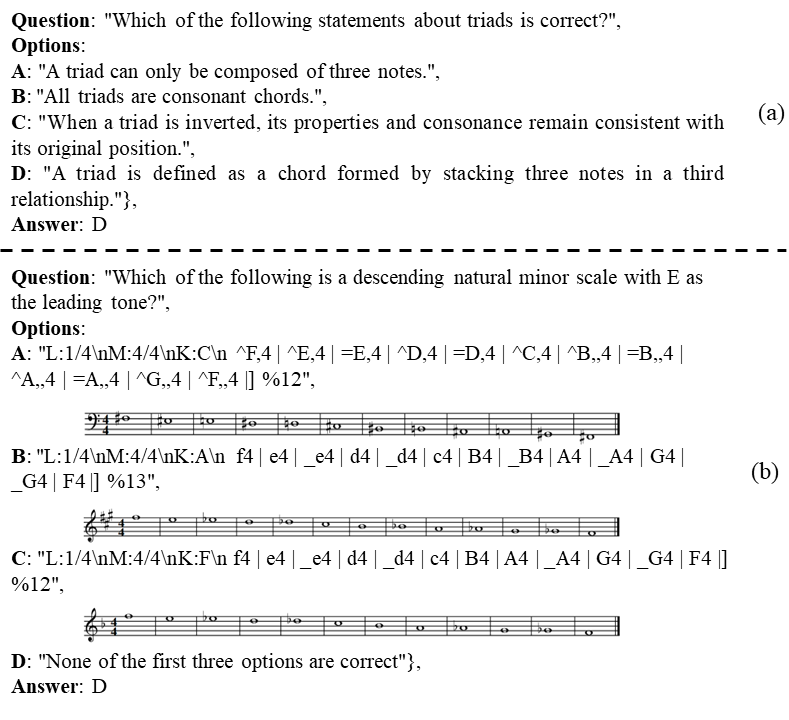}
\end{mdframed}
\caption{Simple examples of (a) music knowledge and (b) music reasoning from MusicTheoryBench. Question a. mainly includes concepts that can be answered through memorizing them. Question b. requires the knowledge of \textit{descending}, \textit{natural minor scale} and \textit{leading tone}, and inference based on the musical score.}
\label{fig:know-reason}
\end{figure}

We define \textit{Music Reasoning} as the capacity to infer the varying harmonies, keys, rhythms, and other musical elements that, although not explicitly annotated in a musical piece, are crucial for understanding its themes, progression, and styles. \textit{Music knowledge}, on the other hand, is defined as the accumulated understanding of musical commonsense, e.g. notions in music theory, history, instrument characteristics, and cultural context, which informs the analytical and creative processes involved in music composition, performance, and appreciation. Examples can be found in Figure \ref{fig:know-reason}.

\paragraph{Curation Process.} We hired a professional college music teacher to craft MusicTheoryBenchmark according to college-level textbooks and exam papers, to ensure consistency with human testing standards. The content underwent multiple rounds of discussions and reviews by a team of musicians. The team carefully selected questions and manually compiled them into JSON and ABC string format. The questions are then labeled into music knowledge and music reasoning subsets. Since the teacher is from China, half of the questions are delivered in Chinese, and later translated into English with GPT-4 Azure API and proofread by the team. 

The resulting benchmark consists of 372 questions, formatted as multiple-choice questions, each with 4 options, among which only one is correct. There are 269 questions on music knowledge and 98 questions on music reasoning, along with 5 questions held out for enabling few-shot evaluation.

\paragraph{Knowledge Subset.} In the music knowledge subset, the questions cover elements from Eastern and Western music. It includes 30 topics such as notes, rhythm, beats, chords, counterpoint, orchestration and instrumentation, music-related culture, history, etc (see \autoref{ref:music knowledge}). Each major area undergoes targeted examination under the guidance of experts and is divided into various subcategories. For example, in the triads section, the test set specifically examines the definition, types, and related technical details of triads. This test also features different levels of difficulty, corresponding to the high school and college levels of music major students.

\paragraph{Reasoning Subset.} Most of the questions in the reasoning subset require both music knowledge and reasoning capabilities. Correctly answering these questions requires detailed analysis of the given information and multi-step logical reasoning, calculating chords, melodies, scales, rhythms, etc.

%% file: tables/pretrain_dataset.tex
\begin{table*}
\centering
\small
\resizebox{\linewidth}{!}{%
\begin{tabular}{llllll}
\toprule
\textbf{Datasets}	& \textbf{Sourced from}		& \textbf{Tokens}	& \textbf{\# Samples}	& \textbf{Category}	& \textbf{Format}  \\
\midrule
Pile \cite{gao2020pile} 								& public dataset								& 0.83B 	& 18K 	& general		& article \\
Falcon-RefinedWeb \cite{penedo2023refinedweb}			& public dataset								& 0.80B 	& 101K 	& general 		& article \\
Wikipedia \cite{2023wiki} 								& public dataset								& 0.39B 	& 588K 	& general 		& article \\
OpenChat \cite{wang2023openchat} 						& public dataset 								& 62.44M 	& 43K 	& general 		& chat \\
LinkSoul \cite{2023linksoul} 						    & public dataset 								& 0.6B 	    & 1.5M 	& general 		& chat \\
GPT4-Alpaca \cite{peng2023instruction} 					& public dataset 								& 9.77M 	& 49K 	& general 		& chat \\
Dolly \cite{DatabricksBlog2023DollyV2} 					& public dataset 								& 3.12M 	& 14K 	& general 		& chat \\
Irishman \cite{wu2023tunesformer} 						& public dataset + Human-written Instructions 	& 0.23B 	& 868K 	& music score 	& chat \\
KernScores \cite{2023kernscores} 						& public dataset + Human-written Instructions 	& 2.76M 	& 10K 	& music score 	& chat \\
Bach \cite{wu2023chord} 								& public dataset + Human-written Instructions 	& 0.44M 	& 349 	& music score 	& chat \\
synthetic music chat\FiveStarCenterOpen 				& public dataset + Human-written Instructions 	& 0.54B 	& 50K 	& music score 	& chat \\
music knowledge\FiveStarConvex 							& Generated w/ GPT-4 							& 0.22B 	& 255K 	& music verbal 	& chat \\
music summary\FiveStarConvex 							& Generated w/ GPT-4 							& 0.21B 	& 500K 	& music verbal 	& chat \\
GSM8k \cite{cobbe2021gsm8k} 							& public dataset 								& 1.68M 	& 7K 	& math 			& chat \\
math \cite{arxiv-math-instruct-50k} 					& public dataset 								& 7.03M 	& 37K 	& math 			& chat \\
MathInstruct \cite{yue2023mammoth} 						& public dataset 								& 55.50M 	& 188K 	& math 			& chat \\
Camel-Math \cite{li2023camel} 							& public dataset 								& 27.76M 	& 50K 	& math 			& chat \\
arxiv-math-instruct-50k \cite{arxiv-math-instruct-50k}  & public dataset 								& 9.06M 	& 50K 	& math 			& chat \\
Camel-Code \cite{li2023camel} 							& public dataset 								& 0.13B 	& 366K 	& code 			& chat \\
OpenCoder \cite{wang2023openchat} 						& public dataset 								& 36.99M 	& 28K 	& code 			& chat \\
\midrule
Total & & 4.16B & 5.17M & & \\
\bottomrule
\end{tabular}
}

\caption{Overview of \dataset. \FiveStarCenterOpen means synthesis from music score data and general data. \FiveStarConvex means with NEW rationales curated by us by prompting GPT-4.}

\label{tab:pretrain_dataset}
\end{table*}

%% file: tables/tasks.tex
\begin{table*}
\Large
\centering
\renewcommand{\arraystretch}{1.2}
\resizebox{\linewidth}{!}{%
\begin{tabular}{@{}lcl@{}}
\toprule
\textbf{Task Name}                                                       & \textbf{Type} & \textbf{Example Instruction}                                                                                                          \\ \midrule
Chord Conditioned Music Generation                              & G         & Develop a musical piece using the given chord progression. {[}CHORDS{]}                                                          \\
Musical Form Conditioned Music Generation                       & G         & Craft a musical work that incorporates the given musical pattern as a central element. {[}MUSICAL FORMS{]}                       \\
Alphabetic Musical Form and Motif Conditioned Music Generation  & G        & Develop a musical piece employing the provided motif and an alphabet-based structure. {[}MUSICAL FORMS A{]} {[}MOTIF{]}          \\
Terminology Musical Form and Motif conditioned Music Generation & G         & Create tunes by incorporating the provided motif in the specified composition structure.  {[}MUSICAL FORMS T{]} {[}MOTIF{]}      \\
Melody Harmonization                                                     & G         & Formulate chord combinations to increase the harmonic complexity of the specified musical excerpt. {[}MELODY{]}                  \\
Bach's Style Music Generation                                   & G        & Provide a musical piece that draws inspiration from Bach's compositions.                                                         \\
Motif Extraction                                                         & U     & Analyze the musical work and pinpoint the consistent melodic element in every section. {[}MUSIC{]}                               \\
Musical Form Extraction                                                  & U      & Investigate the attributes of this musical creation and identify its arrangement using suitable music-related terms. {[}MUSIC{]} \\ \bottomrule
\end{tabular}
}

\caption{Handcrafted musical tasks in \texttt{MusicPile}, including 6 generation tasks (Type:G) and 2 understanding tasks (Type:U), and provide an example prompt for each task. In the examples, we use tokens in square brackets to represent information other than natural language instruction ([MUSICAL FORM A] represents musical form in alphabets and [MUSICAL FORM T] represents musical form in terminology. [MOTIF], [MUSIC] and [MELODY] are represented in ABC notation. [CHORD] is represented in chord symbols.)}
\label{tab:tasks}
\end{table*}

%% file: sections/5experiment.tex
\section{Experiments}

\subsection{Training Settings}
% what model
% what hyperparameter
% what dataset
We initialized a fp16-precision ChatMusician-Base from the LLaMA2-7B-Base weights~\cite{touvron2023llama1, touvron2023llama2}, and applied a continual pre-training plus fine-tuning pipeline. The data settings will be introduced later. 
% The model was trained with fp16 precision. 
LoRA adapters~\cite{hu2021lora} were integrated into the attention and MLP layers, with additional training on embeddings and all linear layers. The maximum sequence length was 2048. We utilized 16 80GB-A800 GPUs for one epoch pre-training and two epoch fine-tuning. DeepSpeed~\cite{rasley2020deepspeed} was employed for memory efficiency, and the AdamW optimizer was used with a 1e-4 learning rate and a 5\% warm-up cosine scheduler. Gradient clipping was set at 1.0. The LoRA parameters dimension, alpha, and dropout were set to 64, 16, and 0.1, with a batch size of 8.
% 16 GPUs 80G A800
% \textbf{Continual Pretraining Setup}

% \textbf{Supervised Finetuning Setup}

\subsection{Data Settings}
% need a table in the appendix about the data mixture settings @hanfeng

During the pretraining, we combined all training data in Section \ref{sec3} and performed one epoch training. 
To explore the effect of different data on the pre-trained model, in the supervised finetuning, we investigated different ratios of data, and empirically determined a 2:1 ratio between music scores and music knowledge\&music summary data. We found that this ratio performed excellently in music generation as well as music understanding while guaranteeing a good MMLU performance. According to the 2:1 ratio, we first sampled 78K samples from the training set and trained for 10 epochs. Then, we maintained the ratio and utilized all available music scores data, which includes 1.1M samples, and trained for 2 epochs. The data mixture settings are summarized in Appendix \ref{data_mixture}.

% overview of data mixture exps

% and reasons for the data mixture exps

\subsection{Evaluation and Baseline Systems}

% gpt4, gpt3.5, llama2
\noindent\textbf{Baseline Systems.} There are currently few LLMs with capabilities in symbolic music. However, observations from \cite{bubeck2023sparks} suggest that the ChatGPT series possesses musical abilities. Therefore, we selected several popular LLM systems, including GPT-3.5, GPT-4, and LLaMA-2, as our baselines.

\noindent\textbf{Evaluation of General Language Abilities.} In order to evaluate general language abilities, we adopt the Massive Multitask Language Understanding (MMLU) dataset \cite{hendrycks2020measuring}, a pioneering benchmark designed to evaluate the knowledge acquired during pretraining of language models. To achieve a fair comparison, we evaluate our models under a 5-shot setting, which keeps the same as our selected baselines.

% \noindent\textbf{Evaluation of Code Abilities.} We evaluate the coding abilities on HumanEval \cite{chen2021evaluating}, an evaluation harness designed to measure functional correctness for synthesizing programs from docstrings. It comprises 164 hand-crafted programming challenges, each including a function signature, docstring, body, and several unit tests, averaging 7.7 tests per problem. Here we report pass@1 metric, which indicates total fraction of problems solved.

% \noindent\textbf{Evaluation of Math Abilities.} We choose GSM8K dataset \cite{cobbe2021training}, a comprehensive collection of 8,500 high-quality, linguistically diverse grade school math problems to evaluate math abilities of our model. This dataset is divided into 7,500 training problems and 1,000 test problems. Each problem in GSM8K necessitates between 2 to 8 steps for resolution, primarily involving basic arithmetic operations such as addition, subtraction, multiplication, and division to arrive at the final answer. These problems are designed in such a way that a bright middle school student should be capable of solving each one, emphasizing the dataset's focus on multi-step mathematical reasoning. We report the 4-shot accuracy on GSM8K.

\noindent\textbf{Evaluation of Music Understanding Abilities.} 
As introduced in Section \ref{music_benchmark}, MusicTheoryBench is a music benchmark proposed in this paper, aiming to inspect the understanding and reasoning capabilities over music knowledge for LLMs. For the MusicTheoryBench, we report the average accuracy after shuffling the option five times as the final results under a zero-shot setting.

\noindent\textbf{Evaluation of Music Generation Abilities.} % subjective metric and case study
Our evaluation of musicality primarily depends on human judgment. Additionally, we have developed two specific metrics: a phrase-level repetition metric and a parsing success rate metric, aimed at assessing the structureness and format accuracy of the generated music. Furthermore, we introduce an average percentile score metric to gauge the models' controllability.

%% file: sections/6results.tex
\section{Results and Discussion}

% qualitative sample
% form control

% quantitative results
% e.g. generate scores that has repeat symbol
% subjective eval
% predict musical form

\subsection{Music Understanding}
\begin{figure}[hbt]
\vspace{0.1cm}
\centering
\setlength{\abovecaptionskip}{0.1cm} 
\includegraphics[width=1.0\columnwidth]{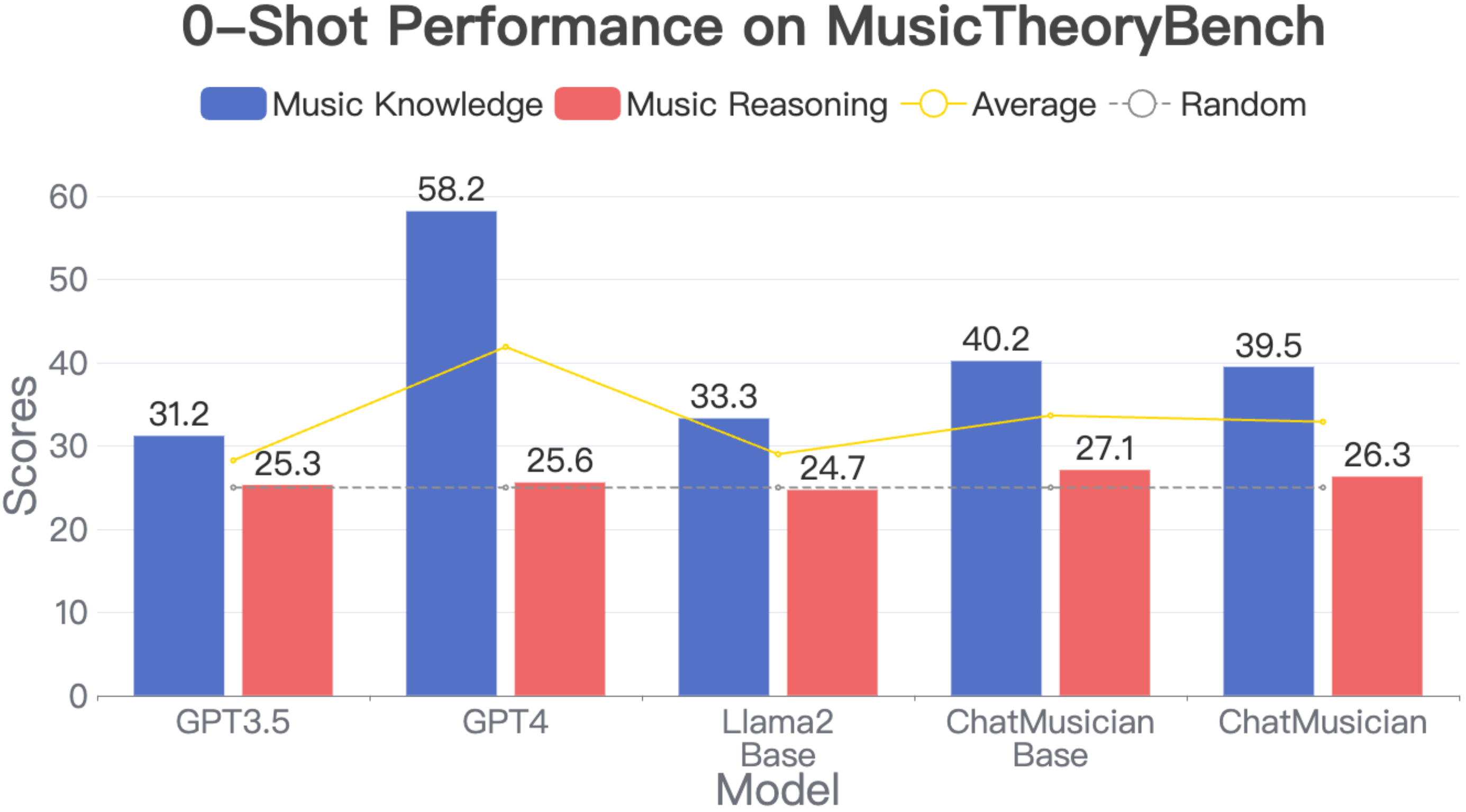}
\caption{Zero-shot accuracy on MusicTheoryBench. We included GPT-3.5, GPT-4, LLaMA2-7B-Base, ChatMusician-Base, and ChatMusician. The blue bar represents the performance on the music knowledge metric, and the red bar represents the music reasoning metric. The dashed line corresponds to a random baseline, with a score of 25\%.
% We use perplexity mode, i.e. selecting the answer with the lowest perplexity, when evaluating LLaMA2 and ChatMusicians.
}
\label{fig:musictheorybenchmark}
\end{figure}

% music knowledge
We use the proposed MusicTheoryBench to evaluate our model and the baseline systems' music understanding abilities. We report the zero-shot performance of GPT3.5, GPT4, LLaMA2-7B-Base, ChatMusician-Base, and ChatMusician on MusicTheoryBench, as shown in Figure~\ref{fig:musictheorybenchmark}. The blue bar represents the performance on the music knowledge metric, and the red bar represents the music reasoning metric. A random baseline corresponds to a score of 25\%, denoted as a dashed line.

\paragraph{Music Knowledge.} According to Figure~\ref{fig:musictheorybenchmark}, all systems significantly surpassed the random baseline in the music knowledge metric. 
GPT-4 achieved the highest score of 58.2 on this metric. Following closely were ChatMusician-Base and ChatMusician, with scores of 40.2 and 39.5, respectively, surpassing GPT-3.5's score of 31.2 and LLaMA2-7B-Base's score of 33.3. This demonstrates the superiority of our method, which significantly enhanced the model's music knowledge capability by around 7 percentage points compared to LLaMA2-7B-Base through continued training. Simultaneously, we observed the alignment tax\cite{zhao2023survey}, where the fine-tuned ChatMusician scored approximately 0.7 points lower on this metric than the Base model.

\paragraph{Music Reasoning.} Contrary to the performance in knowledge metrics, as shown in Figure~\ref{fig:musictheorybenchmark}, all systems exhibit subpar results in music reasoning metrics. The majority of systems do not significantly surpass the baseline in a zero-shot setting. Remarkably, even the most advanced system, GPT-4, only scored 25.6 on this metric. Interestingly, ChatMusician-Base achieved a score of 27.1 in music reasoning metrics, surpassing GPT-4. Furthermore, despite the alignment tax, ChatMusician still obtained a score of 26.3, outperforming GPT-4 in the zero-shot music reasoning metric.

\input{tables/gpt4-musictheorybenchmark}
\paragraph{How Far Can GPT-4 Go?}  MusicTheoryBench represents the first initiative aimed at quantitatively assessing music knowledge and reasoning abilities. In pursuit of this objective, we endeavored to explore the limits of GPT-4 within our benchmark to ascertain its capabilities. GPT-4 is renowned for its robust in-context learning (ICL) and chain-of-thought (CoT) skills. Accordingly, we opted to employ prompt engineering techniques on the GPT-4 baseline to evaluate its performance on the MusicTheoryBench across various conditions, including 5-shots, CoT, and musician role-play prompts.

Table \ref{tab:gpt4-benchmark} displays GPT-4's performance scores under different prompt engineering strategies. Utilizing a combination of role-play and 5-shot ICL techniques, we achieved a peak score of 39.5 in music reasoning. Meanwhile, the integration of CoT and 5-shot ICL techniques resulted in a top score of 69.9 in music knowledge. These results significantly surpass the performance of the vanilla zero-shot approach, yet they still fall short of fully saturating the proposed benchmark.

\subsection{Music Generation}
In this section, we demonstrate that the ABC notation system we have selected serves as an efficient means to encode and compress musical structures and repetitions in a string format. We then provide both qualitative and quantitative evidence to show that our methodology significantly enhances musicality. Moreover, it seamlessly integrates up to six conditional music generation tasks into an LLM without any detrimental effects.

% \vspace{-0.2cm}
% \subsection{Language Ability}
% We report the MMLU score of ChatMusicians, as compared to LLaMA2-7B-Base, in Table~\ref{tab:lang_score}. Our findings indicate that both ChatMusician and ChatMusician-Base achieve higher scores on the MMLU than the LLaMA2-7B-Base model. This suggests that incorporating our method, which infuses intrinsic music understanding and generation capabilities, does not compromise the general language abilities of the model. On the contrary, it appears to enhance them to a certain extent.
% \input{tables/language_score}

\subsubsection{Compression Ratio of ABC Notation}
We sampled a set of 1,000 songs from our training corpus to evaluate the compression ratio of different music representations. As ABC notation can be converted to MIDI or rendered into WAV, we then represent these songs using widely adopted music representations such as ABC strings, MIDI-like, REMI, and audio codecs. We show that the sequence length represented by ABC strings is the shortest, significantly less than other representations. 

As shown in Table \ref{tab:tokens_1k_songs}, ABC notation reaches 288.21 average tokens per song, and 5.16 average tokens per second. This is around 38\% of MIDI-based representations. This suggests that using ABC notation not only facilitates compatibility with text but also reduces training costs and learning complexity.
\input{tables/compression_ratio}

\paragraph{How Does ABC Notation Achieve Such a High Compression Ratio?} The underlying reason is straightforward. The musical score, ingeniously devised by humans, inherently encodes musical repetition. With just a repeat sign denoted as \text{|:} and \text{:|}, repeating phrases or even entire sections can be succinctly notated, corresponding to durations ranging from several seconds to minutes.

\begin{figure}[H]
\vspace{0.1cm}
\centering
\setlength{\abovecaptionskip}{0.1cm} 
\includegraphics[width=.8\columnwidth]{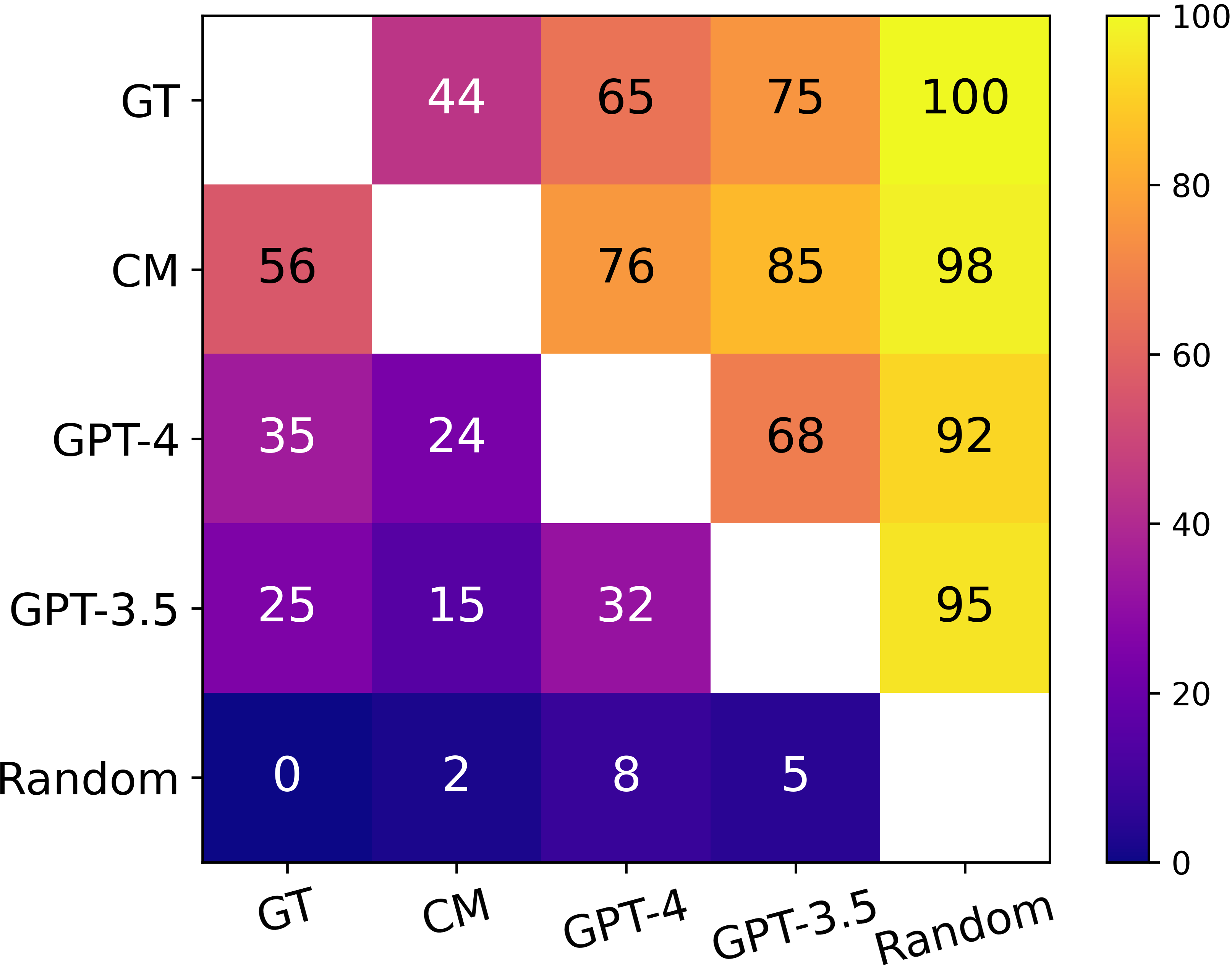}
\caption{Results from our qualitative study where listeners judged pairs of music come from two different sources. Each row indicates the \% of times listeners preferred instrumentals from that system compared to those from each system individually (N = 80). ChatMusician is denoted by CM. i.e.\texttt{76} means that listeners preferred ChatMusician over GPT-4 in 76\% of cases.}
\label{fig:ABtest_matrix}
\end{figure}

\subsubsection{Musicality}
\paragraph{Learning Music Repetitions.} Automatically detecting music repetition and structure remains an unresolved issue. However, by employing ABC notation, we have designed a straightforward experiment to detect phrase-level music repetition by checking the existence of repeat signs. Table~\ref{tab:repetition_det} reports results. It appears that 76\% of the generated samples from ChatMusician contain repeat signs, higher than GPT-4 and GPT-3.5. This suggests that ChatMusician is more likely to generate music with repetition and structure.
\input{tables/repetition_det}

\begin{figure*}[!t]
\begin{mdframed}
% \setlength\multicolsep{0pt}
% \ttfamily\small
% \textbf{Prompt:} Develop melodies by fusing the assigned musical pattern with the given motif. \\
% Binary, Sectional: Verse/Chorus
% \begin{multicols}{2}
% \noindent

\ttfamily\small
\centering
\begin{BVerbatim}[commandchars=\\\{\}]
X:1
L:1/8
M:2/4
K:F
F/G/ \textcolor{blue}{\textbf{|:}}"F" BA"C7" GG |"F" FA"C7" G2 |"F" F>G"C7" AB |
"Am" cA"C7" GF/G/ |"F" BA"C7" GG |"F" FA"C7" G2 |"F" F>G"Bb" Bd |
1"C7" cE"F" FF/G/ \textcolor{blue}{\textbf{:|}}2"C7" cE"F" F z \textcolor{blue}{\textbf{|:}}"F" f3 (c/d/)(d/e/) |
"Gm" (e/f/)(f/g/) g>ec |"C7" e/d/ d/c/c/B/ B/A/A/G/ |"F" GA/B/ c/d/e/f/ | f3 (c/d/)(d/e/) |
"Gm" (e/f/)(f/g/) g>ec |"C7" e/d/ d/c/c/B/ B/A/A/G/ |"F" FA/c/ f z \textcolor{blue}{\textbf{:|}}
\end{BVerbatim}
\par\noindent\rule{\textwidth}{0.4pt}
\centering
\includegraphics[width=.8\textwidth]{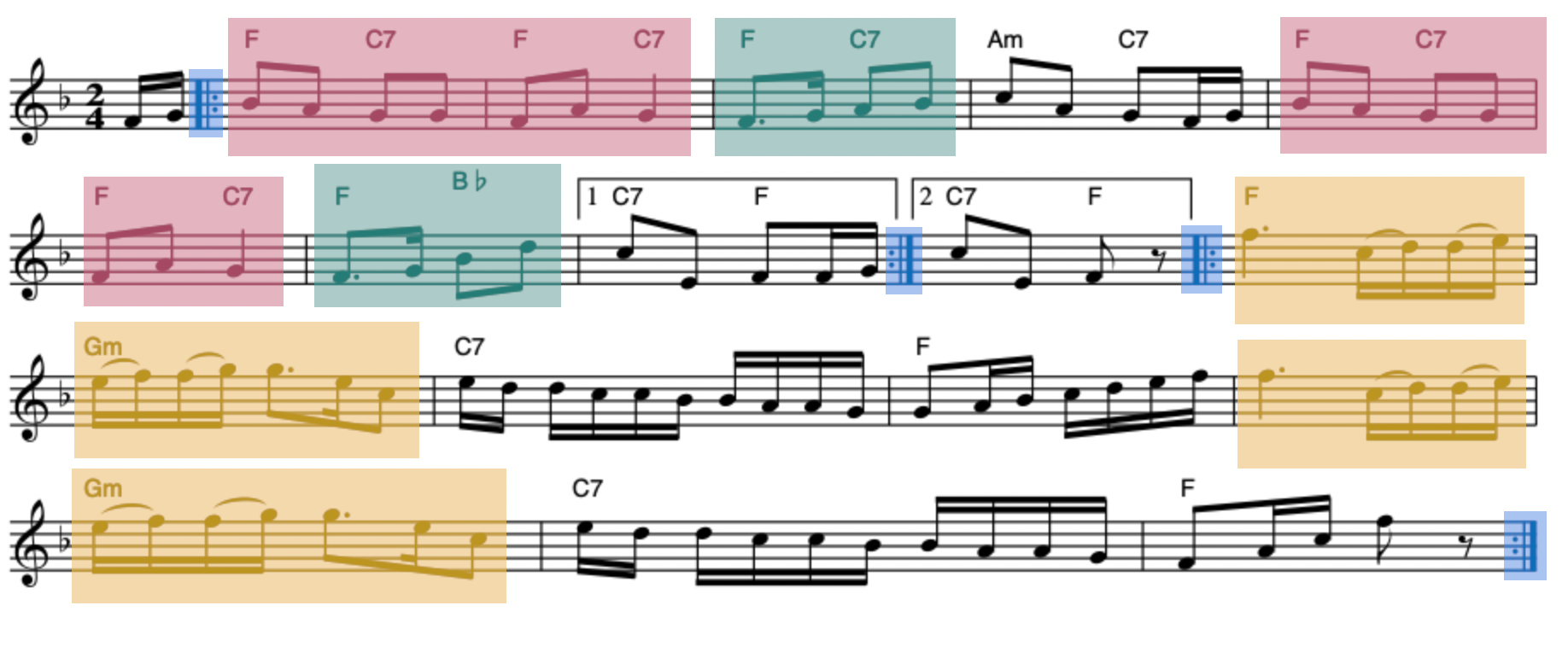}
\caption{ABC notation and corresponding staff notation of a generated music. Repetition symbols are marked blue in both notations and demonstrate a clear phrase-level repetition. Red and yellow rectangles mark clear motif-level repetition in both sections. Green rectangles mark variation notes following the motif of the first section.}
\label{fig:generation_sample}

\end{mdframed}
\end{figure*}

\paragraph{Human Evaluation.}
A more comprehensive evaluation of music repetition and structure requires human assessment. Following \cite{donahue2023singsong} and \cite{thickstun2023anticipatory}, we conduct a listening study to measure the qualitative performance of ChatMusician (\textit{CM}) setting against the ground truth (\textit{GT}) and baselines consisting of \textit{GPT-4}, \textit{GPT-3.5} and random note sequences (\textit{Random}). For our study, listeners are presented with a pair of music excerpts generated from different sources, are asked to indicate which of the two pieces of music excerpts is more musical and are encouraged to pay attention to the musicality from these two aspects: how consistent the music sounds as a whole (e.g., in terms of its melodic contours, rhythmic patterns, and chord progression); and how likely the development of the music follows a clear structure (e.g. verse-chorus division, repetitions). 

Results for all systems appear in Figure~\ref{fig:ABtest_matrix}. When comparing our ChatMusician to GPT-4, listeners preferred music from our system in 76\% of cases. A Wilcoxon signed-rank test of these pairwise judgments indicates that listeners preferred music from CM significantly more often than GPT-4 and GPT-3.5 ($p = 2.7\times10^{-6}$ and $p = 3.8\times10^{-10}$, respectively). 

\paragraph{Qualitative Study} Figure~\ref{fig:generation_sample} presents an example of a generated musical score. The upper area displays the ABC notation, while the lower area illustrates the corresponding staff notation. Notably, repeat signs are marked in blue. Within the phrases, there is also evidence of repetition and motifs, as well as variations that echo these motifs marked in color bars.

% \paragraph{Qualitative Study.} Figure~\ref{fig:generation_sample} presents an example of a generated musical score. Figure~\ref{fig:generation_sample} upper section displays the ABC notation, while Figure~\ref{fig:generation_sample} lower section illustrates the corresponding staff notation. Notably, repeat signs are marked in blue. Within the phrases, there is also evidence of repetition and motifs, as well as variations that echo these motifs, marked in colour bars.

\subsubsection{Controllability}
\paragraph{Format Correctness Evaluation.} We conducted a randomized sampling of 500 music generation prompts from the dataset. These prompts were then processed by ChatMusician, GPT-3.5, and GPT-4 to assess the success rate at which the outputted ABC notation was correctly formatted and parseable. To improve the parsing success rates for the GPT series, we prefixed the prompts with the directive "\textit{Please respond in ABC notation.}". Table ~\ref{tab:format_correctness} presents the comparative success rates across the three systems. Notably, both ChatMusician and GPT-4 demonstrated success rates exceeding 90\%, whereas GPT-3.5 achieved a markedly lower rate of 65.4\%.

\input{tables/format_correctness}
% text, chords, melody, motif, structure, etc.
\paragraph{Task-wise Metrics.} 
We sampled 100 prompts from each of the 5 generation tasks, and calculated average percentile scores as metrics for the 5 music generation tasks, the higher the better. Figure~\ref{fig:subjective_score} presents the detailed score for each task of each model we have tested. See Appendix~\ref{append:average percentile score} for details.
We can see that ChatMusician outperforms both GPT-3.5 and GPT-4 at all five tasks. Note that the low score of GPT-4 at task "Alphabetic musical form and motif music generation" is because most samples generated by GPT-4 of this task contain malformed ABC notation.

\begin{figure}[!t]
\vspace{0.1cm}
\centering
\setlength{\abovecaptionskip}{0.1cm} 
\includegraphics[width=\columnwidth]{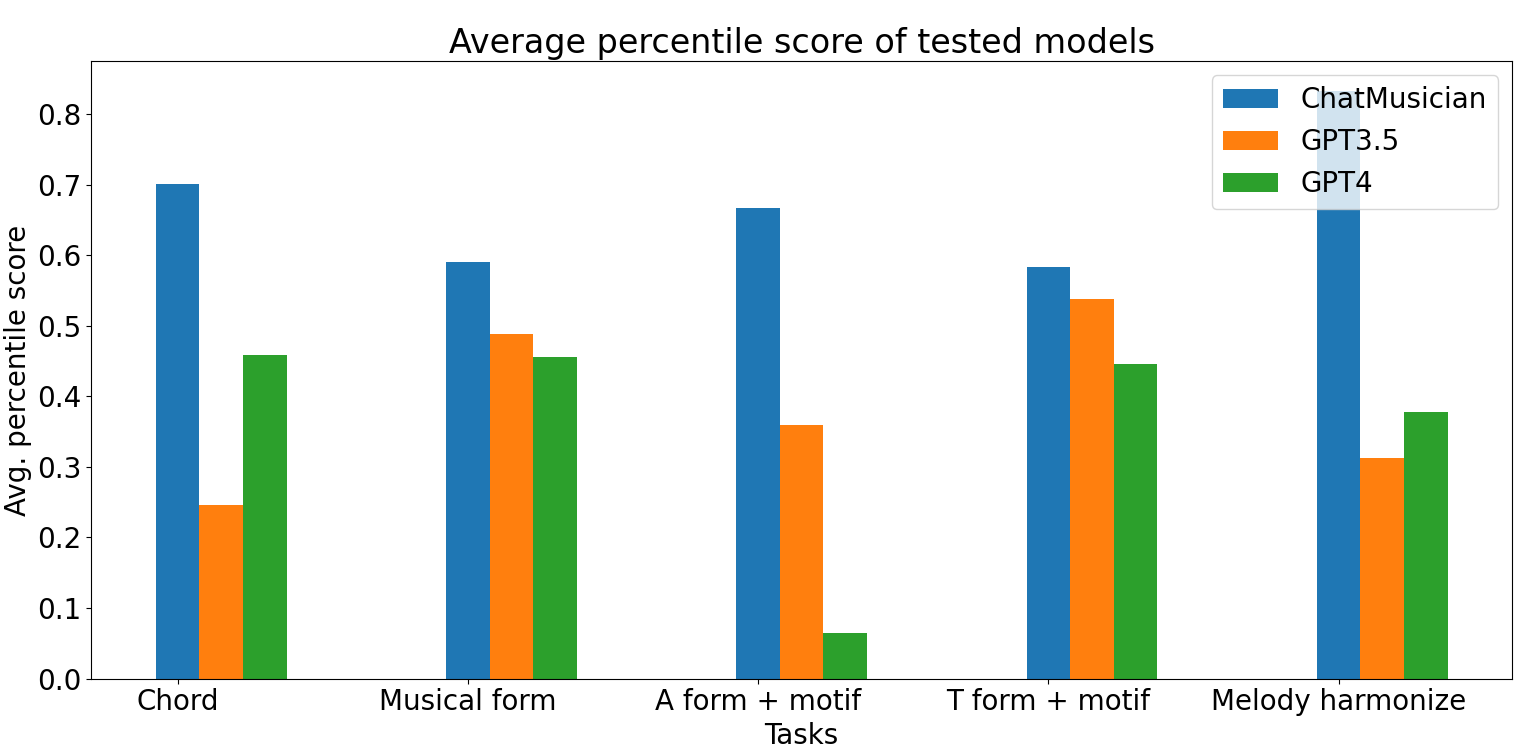}
\caption{Here we provide the average percentile score for 5 out of 8 total musical tasks of ChatMusician, GPT-3.5 and GPT-4. Task names are abbreviations of the tasks in  Table~\ref{tab:tasks} (A form + motif is the abbreviation for "Alphabetic musical form and motif conditioned music generation" and T form + motif is the abbreviation for "Terminology musical form and motif conditioned music generation").}
\label{fig:subjective_score}
\end{figure}

% \paragraph{How Does Learning Music Affects Language Related Abilities?}

% \paragraph{Relationship Between Music Generation and Music Understanding.}

% \subsection{Generation Diversity of ChatMusician}

\vspace{-0.2cm}
\subsection{Language Ability}
We report the MMLU score of ChatMusicians, as compared to LLaMA2-7B-Base, in Table~\ref{tab:lang_score}. Our findings indicate that both ChatMusician and ChatMusician-Base achieve higher scores on the MMLU than the LLaMA2-7B-Base model. This suggests that incorporating our method, which infuses intrinsic music understanding and generation capabilities, does not compromise the general language abilities of the model. On the contrary, it appears to enhance them to a certain extent.
\input{tables/language_score}

\vspace{-0.2cm}
\subsection{Memorization Effect of ChatMusician}
% low priority for first draft
We analyze the memorization abilities of ChatMusician following \cite{copet2023simple}. We randomly select 500 samples from our training set and we feed the model with an instruction prompt. We compare the generated ABC notations with the ground truth. The fraction of examples where the generated and ground truth tokens are identical for the entire sequence is 0.02\%. Furthermore, partial matches occur in 0.24\% of the training examples, where the generated and ground truth sequences share at least 80\% of their tokens.

% does num of epoch affects abilities
% best, last, last3 avg
% 通用data对music knowledge或reasoning或综合music的影响
% music对通用、code、math的影响
% music knowledge和music reasoning两个能力间的关系
% music内部的配比（14,15,16,18,22）
% music generation和music understanding的关系
% 数据量的影响

% zero-shot折叠，5-shot用的人多

% 5 6 7 8的math和code都越训越低
% math和code在5678上相关性大
% music reasoning 的 last3avg 的 20 号实验看出来，code能促进music reasoning
% 

% sft_ablation_base_pt_epoch-1-step-390000 是data mixture的base

%% file: tables/gpt4-musictheorybenchmark.tex
\begin{table}[!t]
    \centering
    \scalebox{0.85}{
    \begin{tabular}{lcc}
        \hline
        Method & Mus. Knowledge  & Mus. Reasoning \\
        % cmidrule
        \hline
        GPT4-0-shot             & 58.2  & 25.6 \\
        \cellcolor{mygray-bg}\quad +5-shot ICL & \cellcolor{mygray-bg}64.1 & \cellcolor{mygray-bg}38.0\\
        GPT4-RolePlay           & 68.3 & 36.6 \\
        \cellcolor{mygray-bg}\quad +5-shot ICL & \cellcolor{mygray-bg}68.8 & \cellcolor{mygray-bg}\textbf{39.5}\\
        GPT4-CoT                & 68.4 & 36.7 \\
        \cellcolor{mygray-bg}\quad +5-shot ICL & \cellcolor{mygray-bg}\textbf{69.9} & \cellcolor{mygray-bg}34.9\\
        \hline
    \end{tabular}
    }
    \caption{We further conducted prompt engineering on GPT-4 to check the upper limit on MusicTheoryBench. We included the techniques of chain-of-thoughts, role-play, and 5-shot in-context-learning. The highest score we achieved on music knowledge metric is 69.9, and 39.5 on music reasoning metric.}
    \label{tab:gpt4-benchmark}
    % \vspace{-10mm}
\end{table}

%% file: tables/compression_ratio.tex
\begin{table}[H]
\centering
\small
%\resizebox{\linewidth}{!}{
\scalebox{0.85}{
\begin{tabular}{llllll}
\toprule
\textbf{Format}	&\textbf{Tokenizer}    & \textbf{Tok./Song}   & \textbf{Tok./Sec.}  \\
\midrule
ABC & LLaMA Tokenizer & 288.21 & 5.16 \\
MIDI & REMI\cite{DBLP:journals/corr/abs-2002-00212} & 753.41 & 12.84 \\
MIDI & MIDI-like\cite{DBLP:journals/corr/abs-1808-03715} & 728.60 & 12.42 \\
WAV & EnCodec\cite{défossez2022high} & 12577.46 & 200.00 \\
\bottomrule
\end{tabular}
}

\caption{The average number of tokens per song (Tok./Song) and tokens per second (Tok./Sec) on 1000 songs with different encoding methods. ABC notation achieves the best compression ratio.}
\vspace{-4mm}
\label{tab:tokens_1k_songs}
\end{table}

%% file: tables/repetition_det.tex
\begin{table}[H]
    \centering
    \scalebox{1}{
    \begin{tabular}{lc}
        \hline
        System & Repetition Det. Rate(\%) \\
        % cmidrule
        \hline
        ChatMusician             & \textbf{76.0}   \\
        GPT-4           & 70.2 \\
        GPT-3.5                & 32.2 \\
        \hline
    \end{tabular}
    }
    \caption{We calculate the phrase-level repetition detection rate of ABC notation strings generated by ChatMusician, GPT-4, and GPT-3.5. The higher the better.}
    \label{tab:repetition_det}
    % \vspace{-10mm}
\end{table}

%% file: tables/format_correctness.tex
\begin{table}[hbt]
    \centering
    \scalebox{1}{
    \begin{tabular}{lc}
        \hline
        System & Success Rate(\%) \\
        % cmidrule
        \hline
        ChatMusician             & \textbf{99.6}   \\
        GPT-4           & 94.6 \\
        GPT-3.5                & 65.4 \\
        \hline
    \end{tabular}
    }
    \caption{We evaluated the parsing success rates of ABC notation strings generated by ChatMusician, GPT-4, and GPT-3.5.}
    \label{tab:format_correctness}
    % \vspace{-10mm}
\end{table}

%% file: tables/language_score.tex
\begin{table}[!t]
    \centering
    \scalebox{1}{
    \begin{tabular}{lc}
        \hline
        System & MMLU Score(\%) \\
        % cmidrule
        \hline
        ChatMusician-Base             & \textbf{48.50}   \\
        ChatMusician           & 46.80 \\
        LLaMA2-7B-Base                & 46.79 \\
        \hline
    \end{tabular}
    }
    \caption{MMLU score of ChatMusicians and LLaMA2-7B-Base.}
    \label{tab:lang_score}
    % \vspace{-10mm}
\end{table}

%% file: sections/7conclusion.tex
\vspace{-0.2cm}
\section{Conclusions}
In conclusion, our study introduces ChatMusician, an innovative open-source LLM capable of advanced music reasoning and composition. By leveraging a text-compatible music representation and achieving notable performance on both music and language benchmarks, ChatMusician represents a significant step forward in integrating musical creativity within language models. Our findings underscore the potential of LLMs as powerful tools for music understanding and creativity, highlighting the untapped possibilities in the fusion of music and artificial intelligence. The release of MusicPiles, MusicTheoryBench, and ChatMusician provides a valuable resource for further research in this exciting domain.

%% file: sections/8limitations.tex
\section*{Limitation}
The current iteration of ChatMusician predominantly generates music in the style of Irish music, attributable to a significant portion of the dataset being sourced from this genre. The model exhibits hallucinations and faces limitations in supporting open-ended music generation tasks due to the lack of diversity in handcrafted music instructions.

% \subsection{Unsatisfied Music Understanding Abilities}

% \subsection{Hallucination}
% hallucination is a thing for solving understanding problems

% but is creative process a kind of hallucination?

% \subsection{Limited Open-ended Music Generation Instruction Following Abilities}
% we need more music data

\section*{Ethics Statement}
The model exhibits illusions, which, if employed in music education, could potentially mislead learners. Additionally, the model demonstrates memorization effect, raising concerns about the potential infringement of music copyrights if it inadvertently regurgitates private training data. We plan to develop a music plagiarism detection algorithm to identify instances of the memorization effect. Furthermore, we aim to implement further alignment strategies to mitigate the occurrence of illusions.

%% file: sections/9acknowledgement.tex
\noindent
\textbf{Core~~~~~~~~~~~~~~~} \\
\noindent
Ruibin Yuan, \textit{ryuanab@connect.ust.hk} \\
Hanfeng Lin, \textit{linhanfeng@bjtu.edu.cn} \\
Yi Wang, \textit{ezzmonyi@gmail.com} \\
Zeyue Tian, \textit{ztianad@connect.ust.hk} \\
Shangda Wu, \textit{shangda@mail.ccom.edu.cn} \\
\\
\textbf{Contributors} \\
Tianhao Shen \\
Ge Zhang \\
Yuhang Wu \\
Cong Liu \\
Ziya Zhou \\
Ziyang Ma \\
Liumeng Xue \\
Ziyu Wang \\
Qin Liu \\
Tianyu Zheng \\
Yizhi Li \\
Yinghao Ma \\
Yiming Liang \\
Xiaowei Chi \\
Ruibo Liu \\
Zili Wang \\
Pengfei Li \\
Jingcheng Wu \\
\\
\\
\textbf{Advisors} \\
Chenghua Lin \\
Qifeng Liu \\
Tao Jiang \\
Wenhao Huang \\
Wenhu Chen \\
Emmanouil Benetos \\
Jie Fu \\
Gus Xia \\
Roger Dannenberg \\
\\
\textbf{Correspondence} \\
Wei Xue, \textit{weixue@ust.hk} \\
Shiyin Kang, \textit{shiyin.kang@kunlun-inc.com} \\
Yike Guo, \textit{yikeguo@ust.hk} \\
\\

%% file: sections/appendix.tex
\newpage
\appendix
\section{Introduction to ABC Notation}\label{sec:abc_introducion}

ABC notation is a text-based system for music notation, particularly popular for notating folk and traditional tunes. It was designed to be easily read by humans and to be simple to type without special musical fonts or software. It offers unique advantages when interfacing with deep learning models:

\begin{itemize}
    \item \textbf{Data Efficiency}: ABC notation compactly represents musical information in a text format, making it highly efficient in terms of storage and transmission. This compactness is advantageous when training deep learning models as it minimizes data overhead.
    
    \item \textbf{Easy Preprocessing}: Being a structured text format, ABC notation can be easily tokenized and converted into numerical sequences or embeddings, a crucial step in preparing data for neural networks.
    
    \item \textbf{Scalability}: The simplicity of ABC notation allows for rapid collection and annotation of large datasets. Deep learning models, especially neural networks, benefit immensely from vast datasets, enabling better training and generalization.
    
    \item \textbf{Generative Models}: ABC notation's text-based nature makes it an excellent candidate for generative models like LLMs, which have shown proficiency in generating coherent sequences in text-based domains.
    
    \item \textbf{Interpretability}: The outputs generated by deep learning models using ABC notation are human-readable, allowing for immediate feedback and iterative refinement. This is particularly useful in tasks like music generation where understanding and tweaking the generated output is crucial.
    
    \item \textbf{Integration with Other Modalities}: ABC notation can be easily integrated with other data modalities in multi-modal deep learning systems, offering a comprehensive representation for music-related tasks.
    
    \item \textbf{Community Support}: The vast number of available tunes and compositions in ABC notation means that there's a rich dataset readily available. Deep learning models can leverage this to learn diverse musical structures and styles.
\end{itemize}

An ABC file consists of a series of headers followed by the music notation. The headers provide metadata about the tune, like its title, composer, rhythm, and more. The music notation section defines the melody.

Headers usually begin with a single letter followed by a colon. Some common headers include:

\begin{mdframed}[innerleftmargin=5pt, innerrightmargin=5pt]
\footnotesize
\begin{verbatim}
X: Reference Number
L: Default Note Length
Q: Tempo
M: Time Signature
K: Key Signature
\end{verbatim}
\end{mdframed}

The music is represented using letters, numbers, and symbols:

\begin{itemize}
    \item Notes are denoted by the letters a-g (for notes in the octave above middle C) and A-G (for the octave below).
    \item Note duration is given by appending a number. For instance, \texttt{A2} indicates a note twice the default length.
    \item Sharps, naturals, and flats are shown with \texttt{\^{}}, \texttt{=}, and \texttt{\_}. For example, \texttt{\^{}F} is an F sharp.
    \item Chords are grouped using square brackets, like \texttt{[ceg]} for a C major chord.
    \item Bars are marked by the \texttt{|} symbol.
    \item Tuplets, like triplets, are notated using special syntax, e.g., \texttt{(3abc} for a triplet of a, b, and c.
    \item Various decorations and ornamentations have unique symbols.
\end{itemize}

Here's a basic tune in ABC notation:

\begin{mdframed}[innerleftmargin=5pt, innerrightmargin=5pt]
\footnotesize
\texttt{%
X:1\\
L:1/8\\
M:3/4\\
K:D\\
de |"D" f3 g a{/g}f |"A" e4 AB |"C" =c3 d"G/B" B{/A}G |"A" A4 fg |\\
"D" a3 g fd |"A" e4 AB |"C" =c3 d e{/f}g |"D" f4 ::\\
d{/edc}A |"Bm" B2 A2 F2 |"F\#m" A3 B d{/edc}A |"G" B2 A2 F2 |\\
"A" (E4 E)A |"Bm" B3 A FD |"F\#m" C3 D (F/G/A) |\\
"G" B3 e"A" dc |"D" d4 :|
}
\end{mdframed}

This represents a waltz set in D major. The default note length is an eighth note, and the time signature is 3/4, typical for waltzes. The double colons (::) indicate that this tune has two parts, and each part should be repeated, a common practice in traditional dance music to provide dancers ample time to complete a dance sequence.

\section{The examination scope of MusicTheoryBench}
\label{ref:music knowledge}

\begin{enumerate}
  \item Pitch, Note Value, and Notation System
  \begin{itemize}
      \item Sound, Musical Tone Characteristics
      \item Musical Tonality System, Tone Row, and Scale Degrees
      \item Grouping of Notes
      \item Staff, Clef and Stave
      \item Division of Note Values
      \item Semitone and Whole Tone
      \item Temperament
      \item Harmonic Series
  \end{itemize}
  \item  Rhythm, Beat and Note Value Combinations
  \begin{itemize}
      \item Rhythm and Beat
      \item Even Rhythmic Division and Irregular or Special Rhythmic Division
      \item Time Signature and Types of Time Signatures
      \item Syncopation
      \item Note Value Combination
  \end{itemize}
  \item Interval
  \begin{itemize}
      \item Definition and Classification of Intervals
      \item Degrees and Intervals
      \item Diatonic and Chromatic Intervals
      \item Single Intervals and Compound Intervals
      \item Inversion of Intervals
      \item Consonant Intervals and Dissonant Intervals
      \item Enharmonic Intervals
      
  \end{itemize}
  \item Triad
  \begin{itemize}
      \item Definition of Triad
      \item Types of Triads
      \item Inversions of Triads
  \end{itemize}
  \item Seventh chord
  \begin{itemize}
      \item Definition and Types of Seventh Chords
      \item Positions and Inversions of Seventh Chord
      \item Arrangements of Seventh Chords
      \item Enharmonic Chord
      \item Consonance of Chords
      \item Ninth Chord, Eleventh Chord and Thirteenth Chord
  \end{itemize}
  \item Modal Scales
  \begin{itemize}
      \item Key Name, Key Signature and Scale Degrees
      \item Major Scale
      \item Minor Scale
      \item Medieval Modes
      \item Ethnic Scales
  \end{itemize}
  \item  Relationship between Keys
  \begin{itemize}  
     \item  Relationship between Major and Minor Keys.
      \item Modal Interchange
      \item Relative Major and Minor
      \item Tone Equal Temperament
      \item Relative Keys
  \end{itemize}
  \item Western Modes and Tonality
  \begin{itemize}
      \item Mode and Key Signature
      \item Natural Major and Minor Scales
      \item Harmonic Major and Minor Scales
      \item Melodic Major and Minor Scales
      \item Tonal Chromaticism in Modal Analysis
  \end{itemize}
  \item Ethnic Modal Scales and Tonality
  \begin{itemize}
      \item Pentatonic Scale
      \item Hexatonic Scale
      \item Heptatonic Scale
  \end{itemize}
  \item Transposition
  \begin{itemize}
      \item Western Transposition
      \item Ethnic Transposition
  \end{itemize}
  \item Tonal Analysis of Chord Progressions
  \item Intervals and Chords in a Mode.
  \begin{itemize}
      \item  Intervals in a Mode
      \item Resolution of Intervals
      \item Triads in a Mode
      \item Seventh Chords in a Mode
      \item Resolution of Dominant Seventh Chords
  \end{itemize}
  \item Transposition in Notation
  \begin{itemize}
      \item Interval Transposition
  \end{itemize}
  \item Chromatic Scale
  \begin{itemize}
      \item Major Chromatic Scale
      \item Minor Chromatic Scale
      \item Dynamic Markings and Terminology
      \item Tempo Markings and Terminology
      \item  Inversions and Voicings
      \item Augmented Sixth Chords
      \item Neapolitan and Borrowed Chords
  \end{itemize}
  \item Form and Structure
  \begin{itemize}
      \item Phrases, Periods and Sentences
      \item Binary, Ternary and Rondo Forms
      \item Sonata-Allegro Form
      \item Theme and Variations
      \item Fugue and Other Contrapuntal Forms
  \end{itemize}
  \item Counterpoint 
  \begin{itemize}
      \item Species Counterpoint
      \item The Rules of Voice-Leading
      \item Imitative Counterpoint (Canon, Fugue)
      \item Imitative Counterpoint
  \end{itemize}
    \item Melody 
    \begin{itemize}
        \item Melodic Construction and Development
        \item Motivic Development
        \item Sequences
    \end{itemize}
    \item Twentieth-Century Techniques
    \begin{itemize}
        \item Atonality and Serialism
        \item Twelve-Tone Technique
        \item Set Theory
        \item Minimalism 
        \item Microtonality 
    \end{itemize}
    \item Musical Styles and Genres
    \begin{itemize}
        \item Historical Overview from Medieval to Contemporary
        \item Characteristics of Different Musical Periods (e.g., Baroque, Classical, Romantic)
    \end{itemize}
    \item Analysis Techniques
    \begin{itemize}
        \item Roman Numeral Analysis
        \item Schenkerian Analysis
        \item Graphic Analysis
        \item Neo-Riemannian Theory
    \end{itemize}
    \item Orchestration and Instrumentation.
    \begin{itemize}
        \item Characteristics of Orchestral Instruments
        \item Basics of Writing for Different Instruments
        \item Full Orchestral Scoring
    \end{itemize}
    \item Acoustics and the Science of Sound 
    \begin{itemize}
        \item Overtones and Harmonics
        \item The Harmonic Series
        \item Timbre and Its Characteristics 
    \end{itemize}
\end{enumerate}

\section{Examples used in 5-shot evaluation in MusicTheoryBench}
As shown in Figure \ref{fig:five_shot_examples}, we present our held-out examples with prompt used in 5-shot evaluation.

% \begin{figure}[h]
% \begin{mdframed}[innerleftmargin=1pt, innerrightmargin=1pt]
% \footnotesize
% \includegraphics[width=\linewidth]{figures/fig-know-reason.png}
% \end{mdframed}
% \caption{Simple examples of (a) music knowledge and (b) music reasoning from MusicTheoryBench. Question a. mainly include concepts that can be answered through memorizing them. Question b. requires the knowledge of \textit{descending}, \textit{natural minor scale} and \textit{leading tone}, and inference based on the musical score.}
% \label{fig:know-reason}
% \end{figure}

\begin{figure*}[p]
\begin{mdframed}[innerleftmargin=5pt, innerrightmargin=5pt]
\footnotesize
\texttt{%
Read the following questions from the four options (A, B, C and D) given in each question. Choose the best option.\\
Which of the following chord progressions best describes the above example?\\
L:1/4\\
M:4/4\\
K:E\\
 \texttt{[G,B,E]} [A,CE] [F,B,D] [F,A,C] |] \%1\\
A. ii\\
6\\
/\\
4 – V – vi\\
6 - iii\\
\\
B. I\\
6 – IV – V6\\
/\\
4 - ii\\
C. IV – V6\\
/\\
4 – I - ii\\
\\
D. iii\\
6 – V – I\\
6\\
/\\
4 - IV\\
Answer: B\\
\\
Which of the following best describes the seventh chord in the above example?\\
L:1/4\\
M:4/4\\
K:D\\
 \texttt{[FGBd]4 |]} \%1\\
A. Major seventh in third inversion\\
B. Dominant seventh in second inversion\\
C. Major/minor seventh in third inversion\\
D. Minor seventh in second inversion\\
Answer: A\\
\\
Which of the following is the name of the note in the above example?\\
L:1/4\\
M:4/4\\
K:Cb\\
 \texttt{D,4 |]} \%1\\
A. B-flat\\
B. D\\
C. B\\
D. D-flat\\
Answer: D\\
\\
The chord in the above example can be best described as which of the following?\\
L:1/4\\
M:4/4\\
K:F\#\\
 \texttt{[EGB]4 |]} \%1\\
A. viio\\
B. V\\
C. ii\\
D. iv\\
Answer: A\\
\\
\texttt{[Actual question here]}
}
\end{mdframed}

\caption{5-shot examples and prompt used in MusicTheoryBench benchmark.}
\label{fig:five_shot_examples}
\end{figure*}

\section{Music Instruction Dataset Curation}\label{sec:music_score_set}
We used the Irishman dataset as the basis of our music SFT data. The original dataset contains two
fields: control code and ABC notation. Control code is the instruction to generative model on the overall structure of the generated symbolic music. Here we provide a control code sample for a better explanation: 

\begin{doublespace}
\centerline{\texttt{S:2 B:5 E:5 B:6 S:2}}
\end{doublespace}

S:2 represents that there are 2 sections in this music sample, each section
would be clearly marked by segmentation marks in ABC notation. B:5 represents
that there are 5 bars in the first section, and B:6 represents that the second section contains 6 bars. E:5 between the two B sections represents the edit distance similarity
between two music sections, in this sample: 0.5. 

For the $n^{\text{th}}$ B section, there exists $n-1$ number of E sections before it, in which the $m^{\text{th}}$ E section represents the edit distance similarity between the $m^{\text{th}}$ B section and the $n^{\text{th}}$ B section.
\subsubsection{Musical form analysis algorithm}
For each E section before a B section, we can build a list of similarity levels for the current B section. In each of these lists, we use the following standards: 

Similarity greater than or equal to 8 represents two sections that can be seen as identical sections, notated as \textit{s}. The similarity between 6 and 8 represents a section that can be seen as a variation of the previous sections, notated as \textit{v}. Similarity under 6 represents two different sections, notated as \textit{d}). Give the following example of control code to algorithm 2:

\begin{doublespace}
\centerline{\small{\texttt{S:4 B:1 E:1 B:8 E:3 E:7 B:1 E:1 E:4 E:1 B:8}}}
\end{doublespace}

we would get this similarity level list $a=$ [[\textit{d}], [\textit{d}, \textit{v}], [\textit{d}, \textit{d}, \textit{d}]]

Then we create a string to represent the alphabet musical form and put character \textit{A} at its beginning, walk through each sub-list in the similarity level list, and mark the index of the first appeared \textit{s} and \textit{v}. If \textit{s} $>$ \textit{v}, we will append the same alphabet at the index of \textit{s}. If \textit{v} $>$ \textit{s}, we will append the alphabet at the index of \textit{v} with an added prime sign. 

In the example above, we would get its alphabet musical form as $ABB'C$.

Using this alphabetic musical form, we can produce musical forms represented by terms. We gathered some commonly used musical form terms and put them into three categories: traditional musical forms from music theory, including \textit{Only One Section}, \textit{Binary}, \textit{Ternary}, \textit{Variational}, extended musical forms, including \textit{American Popular}, \textit{Verse/Chorus}, \textit{Verse/Chorus/Bridge}, \textit{Verse/Chorus/Verse/Bridge}, \textit{Through Composed}, and compound musical forms, including \textit{Compound Binary}, \textit{Compound Ternary}.

\subsubsection{Motif extraction algorithm}
The motif extraction algorithm starts by separating the sample into each section with section length information provided in the control code, then processes the token sequence $s$ of each section with algorithm 1.
\begin{algorithm}[ht]
   \caption{ABC Notation Motif Extraction}
   \label{motif_extraction}
\begin{algorithmic}
\State \textbf{Input:} $s^{(0)}\cdots s^{(n)}$
\For{$x=0,1,\cdots,n$}
\If{$s^{(x)}$ is a bar, chord, annotation, decoration symbols, decoration characters, embellish symbols}
Drop $s^{(x)}$ from $s$
\EndIf
\EndFor
\State Suppose $m$ is the new token sequence length. Create an empty token frequency tuple list $a$.
\For{$y=0,1,\cdots,m$} \For{$z=1,2,\cdots,8$} 
\If{$s^{(y)},\cdots,s^{(y+z)} \notin a$} 
\State Add $(s^{(y)},\cdots,s^{(y+z)},1)$ to $a$
\Else{ Get tuple $(s^{(y)},\cdots,s^{(y+z)},k)$ from $a$ and update it to $(s^{(y)},\cdots,s^{(y+z)}, k+1)$}
\EndIf
\EndFor
\EndFor
\State From $a$ get the tuple $b$ with the largest $b[1]$ value and $len(b[0])$ value.
\State \Return $b[0]$ as the motif 

\end{algorithmic}
\end{algorithm}

\begin{algorithm}[ht]
   \caption{Control Code Based Musical Form Analysis}
   \label{musical_form_analysis}
\begin{algorithmic}
\State \textbf{Input:} $s^{(0)}\cdots s^{(n)}$
\State Create musical form string $m=$"A", two empty list $a$ and $b$, current alphabet $n=$ `A'
\For{$x=1,2,\cdots,n$}
\If{$s^{(x)}[0]=$"B"} 
\State Append $b$ to $a$, create a new empty list $b$
\Else 
  \If{$s^{(x)}[-1] \geq 8$} 
  \State Append "s" to $b$
  \ElsIf{$8>s^{(x)}[-1] \geq 6$} 
  \State Append "v" to $b$
  \Else{ Append "d" to $b$}
  \EndIf
\EndIf
\EndFor
\For{sub list $c$ in list $a$}
\State $p_v=c.index$("v") 
\State $p_s=c.index$("s")
\If{$p_v > p_s$} 
\State Append $m$[$p_v$]+' to $m$
\ElsIf{$p_v < p_s$} 
\State Append $m$[$p_s$] to $m$
\Else{ Append current alphabet $n$ to $m$ and move $n$ to the next alphabet}
\EndIf
\EndFor
\State \Return $m$ as the musical form
\end{algorithmic}
\end{algorithm}

\section{Settings of Data Mixture}\label{data_mixture}

\input{tables/sft_exp}
To consider the limited computing power and explore the impact of data mixtures, we downsampled our data to a size of 52k or 78k, with different mixture proportions. This allows for the experiment to be completed in approximately one day.  All the settings are in Table \ref{tab:sft_exp}. 
% We also conduct a data filtering 
% The perplexity metric (ppl) of all samples was calculated using the pre-trained model. Samples with a ppl greater than the median were filtered out and only high-quality samples were selected for experimentation, except for the 21st setting. 
Table \ref{tab:pretrain_dataset} contains the categorization of data domains. Music verbal refers to a combination of music knowledge and music summary. 
% In these settings, we randomly sampled according to the original proportions in their datasets, except for the specially labeled proportions. 
Empirically, we find that setting 18 gives a balanced performance among music understanding, music generation, and language understanding abilities. Subsequently, we scaled up setting 18 to 1.1M samples and denoted it as setting 21. Setting 21 is the reported ChatMusician system in the main paper.

\section{Details of Average Percentile Score Metric}
\label{append:average percentile score}
For each task, we first calculate an initial score. For chord conditioned music generation task, the initial score is calculated by taking the edit distance between the chords in the generated music and the chords in the prompt. For the musical form conditioned music generation task, the initial score is calculated by taking the difference between the set of musical forms calculated from generated music and the set of musical forms in the prompt. For the alphabetic/terminology musical form and motif-conditioned music generation task, the initial score is calculated by taking both the difference between the set of musical forms calculated from generated music and the set of musical forms in the prompt and the longest common sub-sequence of motif calculated from generated music and motif in the prompt. For the melody harmonization task, the initial score is calculated by taking the edit distance between the melody in the generated music and the melody in the prompt.

Since we have different initial score calculation methods for each task, we normalize the score to the same scale by taking the percentile of initial scores under each task. A percentile value represents that the initial value of a sample is larger than how much percentage of all the initial values in this task. For example, a percentile value of 0.6 in chord conditioned music generation task means that the initial score of the sample is larger than 60\% of all the initial scores in chord conditioned music generation task. Finally, we take the average value of the percentile for each task of each model and produce the average percentile score of each task for tested models at Figure~\ref{fig:subjective_score}.

% \begin{figure}[H]
% \vspace{0.1cm}
% \centering
% \setlength{\abovecaptionskip}{0.1cm} 
% \includegraphics[width=\columnwidth]{figures/subjective_score.png}
% \caption{Here we provide the average percentile score for 5 out of 8 total musical tasks of ChatMusician, GPT-3.5 and GPT-4. Task names are abbreviations of the tasks in  Table~\ref{tab:tasks} (A form + motif is the abbreviation for "Alphabetic musical form and motif conditioned music generation" and T form + motif is the abbreviation for "Terminology musical form and motif conditioned music generation").}
% \label{fig:subjective_score_ap}
% \end{figure}

% \begin{figure*}[t]
% \centering
% \includegraphics[width=\textwidth]{figures/representation.pdf}
% \caption[]{Commonly used music representations, including Wav, Codec, MIDI and ABC notation. From left to right, the compression rate gets higher.
% } 
% \label{fig:representation}
% \end{figure*}

%% file: tables/sft_exp.tex
\begin{table*}[!t]
\centering
\small
\resizebox{\linewidth}{!}{%
\begin{tabular}{llllll}
\toprule
\textbf{ID}	&\textbf{Setting}    & \textbf{\# Samples}   & \textbf{Epochs}  \\
\midrule
1 & general + math + code & 52k & 10 \\
2 & music verbal & 52k & 10 \\
3 & music (verbal + score) & 52k & 10 \\
4 & general + music (verbal + score) + code + math & 78k & 10 \\
5 & music (verbal + score) : general = 1:2 & 78k & 10 \\
6 & music (verbal + score) : general = 2:1 & 78k & 10 \\
7 & general + math + music (verbal + score) & 78k & 10 \\
8 & general + code + music (verbal + score) & 78k & 10 \\
9 & general(exclude linksoul) + music (verbal + score) + code + math & 78k & 10 \\
10 & music verbal : general + math + code = 1:2 & 78k & 10 \\
11 & music verbal : general + math + code = 2:1 & 78k & 10 \\
12 & music verbal : general + math + code (en) = 1:2 & 78k & 10 \\
13 & music verbal : general + math + code (en) = 2:1 & 78k & 10 \\
14 & music verbal : irishman = 5:1 & 52k & 10 \\
15 & music verbal : irishman = 1:1 & 52k & 10 \\
16 & music verbal : synthetic music chat = 5:1 & 52k & 10 \\
17 & music verbal : general(en) = 1:1 & 52k & 10 \\
18 & music verbal : music score = 2:1 & 78k & 10 \\
19 & music verbal + math : music score = 2:1 & 78k & 10 \\
20 & music verbal + code : music score = 2:1 & 78k & 10 \\
21 & music verbal : music score = 2:1 & 1.1M & 2 \\
22 & music verbal : bach = 2:1 & 78k & 10 \\
23 & music verbal : music score(half bach) = 2:1 & 78k & 10 \\
24 & music verbal : music score(bach repeat 10) = 2:1 & 78k & 10 \\
% 25 & music verbal : music score + control code + ap = 2:1 & 78k & 10 \\
\bottomrule
\end{tabular}
}

\caption{Settings of Data Mixture in Supervised Finetuning Phase.}

\label{tab:sft_exp}
\end{table*}